\documentclass{amsart}

\newcommand {\supplus}{\mathop{{\supset}\llap{\raise 
0.5pt\hbox{\normalfont\small+}\hskip 0.5pt}}} 

\newcommand {\subplus}{\mathop{{\subset}\llap{\raise 
0.5pt\hbox{\normalfont\small+}\hskip 0.5pt}}}  

\newcommand {\Cee}    {{\mathbb  C}}

\newcommand {\Pee}    {{\mathbb  P}}

\newcommand {\Zee}    {{\mathbb  Z}}

\newcommand {\fa}     {{\mathfrak{a}}}

\newcommand {\fab}    {{\mathfrak{ab}}} 

\newcommand {\fas}    {{\mathfrak{as}}}
\newcommand {\faut}   {{\mathfrak{aut}}} 
\newcommand {\fb}     {{\mathfrak{b}}}
\newcommand {\fc}    {{\mathfrak{c}}}
\newcommand {\fcg}    {{\mathfrak{cg}}}

\newcommand {\fcvect}   {{\mathfrak{cvect}}}
\newcommand {\fder}   {{\mathfrak{der}}}   %

\newcommand {\fdg}    {{\mathfrak{dg}}}
\newcommand {\fd}     {{\mathfrak{d}}}
\newcommand {\fe}     {{\mathfrak{e}}}

\newcommand {\ff}     {{\mathfrak{f}}}

\newcommand {\fg}     {{\mathfrak{g}}}    %
\newcommand {\fgl}    {{\mathfrak{gl}}}  %
\newcommand {\fh}     {{\mathfrak{h}}}
\newcommand {\fhei}   {{\mathfrak{hei}}}
\newcommand {\fii}    {{\mathfrak{i}}}    %
\newcommand {\fk}     {{\mathfrak{k}}}

\newcommand {\fle}    {{\mathfrak{le}}}
\newcommand {\fm}     {{\mathfrak{m}}}

\newcommand {\fo}     {{\mathfrak{o}}}
\newcommand {\fosp}   {{\mathfrak{osp}}}
\newcommand {\fpe}    {{\mathfrak{pe}}}   %

\newcommand {\fpo}    {{\mathfrak{po}}}

\newcommand {\fs}     {{\mathfrak{s}}}

\newcommand {\fsb}    {{\mathfrak{sb}}}

\newcommand {\fsh}    {{\mathfrak{sh}}}

\newcommand {\fsl}    {{\mathfrak{sl}}}
\newcommand {\fsle}   {{\mathfrak{sle}}}
\newcommand {\fsm}    {{\mathfrak{sm}}}
\newcommand {\fsp}    {{\mathfrak{sp}}}
\newcommand {\fspe}   {{\mathfrak{spe}}}
\newcommand {\fspo}   {{\mathfrak{spo}}}

\newcommand {\fsvect} {{\mathfrak{svect}}}

\newcommand {\fu}     {{\mathfrak{u}}}
\newcommand {\fvect}  {{\mathfrak{vect}}}   %

\newcommand {\fx}  {{\mathfrak{x}}}
\newcommand {\fz}     {{\mathfrak{z}}}

\newcommand {\cal} {\mathcal}

\newcommand {\cC}     {{\cal C}}

\newcommand {\cF}     {{\cal F}}

\newcommand {\cL}     {{\cal L}}
\newcommand {\cM}     {{\cal M}}

\def \opname#1#2%
  {\expandafter\newcommand \csname #1\endcsname {{\mathop{#2}\nolimits}}}

\newcommand{\rmname}[1]
  {\expandafter\newcommand \csname #1\endcsname {{\operatorname{#1}}}}

\newcommand{\rmnameii}[2]
  {\expandafter\newcommand \csname #1\endcsname {{\operatorname{#2}}}}

\rmname{act}
\rmname{Ad}
\rmname{Add}
\rmname{ad}
\rmname{Alt}
\rmname{alt}
\rmname{Ann}
\rmname{antidiag}
\rmname{Ber}
\rmname{ber}
\rmname{Br}
\rmname{card}
\rmname{ch}
\rmname{Char}
\rmname{cem}
\rmname{cj}
\rmname{Cliff}
\rmname{cntr}
\rmname{codim}
\rmname{coind}
\rmname{const}
\rmname{col}
\rmname{cork}
\rmname{cpr}
\rmname{diag}
\rmnameii{Div}{div}
\rmname{Def}
\rmname{Der}
\rmname{Dim}
\rmname{End}
\rmname{Even}
\rmname{Ext}
\rmname{gr}
\rmname{Hom}
\rmname{HT}
\rmnameii{Ht}{ht}
\rmname{hwt}
\rmname{Id}
\rmname{id}
\rmname{ind}
\rmname{Ind}
\rmname{Inf}
\rmname{irr}
\rmname{Le}
\rmname{Lie}
\rmname{lwt}
\rmname{mult}
\rmname{Mor}
\rmname{nm}
\rmname{Ob}
\rmname{Odd}
\rmname{Osc}
\rmname{per}
\rmname{Pic}
\rmname{pr}
\rmname{pro}
\rmname{Prime}
\rmname{Proj}
\rmname{prt}
\rmname{pt}
\rmname{Q}
\rmname{qet}
\rmname{qtr}
\rmname{rd}
\rmname{rk}
\rmname{row}
\rmname{Res}
\rmname{salt}
\rmname{Sch}
\rmname{SBr}
\rmname{scalar}
\rmname{Ser}
\rmname{sign}
\rmname{Smbl}
\rmname{spin}
\rmname{ssym}
\rmname{str}
\rmname{st}
\rmname{sgn}
\rmname{sq}
\rmname{symm}
\rmname{supp}
\rmname{Supp}
\rmname{St}
\rmname{Spec}
\rmname{Spm}
\rmname{tr}
\rmname{vpt}
\rmname{weyl}
\rmname{Weyl}
\rmname{Witt}

\opname{vvol}  {{v\hspace{-0.1ex}o\hspace{-0.02ex}l\/}}
\opname{pnt}  {\text{\normalfont pt}}
\opname{Span} {{Span}}
\opname{slim} {\overline{\lim}}
\opname{Vol}  {{V\hspace{-0.55ex}o\hspace{-0.02ex}l\/}}
\opname{QVol} {{Q\hspace{-0.3ex}V\hspace{-0.55ex}o\hspace{-0.02ex}l\/}}
\opname{PoVol}{{P\hspace{-0.35ex}o\hspace{-0.25ex}V\hspace{-0.55ex}o\hspace{-0.02ex}l\/}}
\opname{BVol} {{B\hspace{-0.2ex}V\hspace{-0.55ex}o\hspace{-0.02ex}l\/}}
\opname{Par}  {{P\hspace{-0.3ex}a\hspace{-0.05ex}r\/}}

\rmname{Mat}
\rmname{Bil}
\rmname{Diff}
\rmname{Ker}
\rmname{Herm}
\rmname{Coker}
\rmname{Conn}
\rmname{Covect}
\rmname{Vect}
\rmname{Int}

\rmnameii {IM} {Im}
\rmnameii {RE} {Re}

\opname{Aut} {{A\hspace{-0.2ex}u\hspace{-0.1ex}t\/}}
\opname{GL} {{G\hspace{-0.3ex}L}}
\opname{SL} {{S\hspace{-0.3ex}L}}
\opname{Exp} {{E\hspace{-0.2ex}x\hspace{-0.1ex}p\/}}
\opname{GQ} {{G\hspace{-0.2ex}Q}}
\opname{OSp} {{O\hspace{-0.25ex}S\hspace{-0.15ex}p\/}}
\opname{Out} {{O\hspace{-0.25ex}u\hspace{-0.15ex}t\/}}
\opname{Spp} {{S\hspace{-0.2ex}p\/}}
\opname{SpO} {{S\hspace{-0.2ex}p\hspace{-0.02ex}O\/}}
\opname{Pe} {{P\hspace{-0.25ex}e\/}}
\opname{SPe} {{S\hspace{-0.25ex}P\hspace{-0.25ex}e\/}}
\opname{Spin} {{S\hspace{-0.25ex}p\hspace{-0.05ex}i\hspace{-0.1ex}n\/}}
\opname{Iso} {{I\hspace{-0.25ex}s\hspace{-0.1ex}o\/}}
\opname{SSPe} {{S\hspace{-0.25ex}S\hspace{-0.15ex}P\hspace{-0.25ex}e\/}}
\opname{PeU} {{P\hspace{-0.25ex}e\hspace{-0.1ex}U\/}}
\opname{QU} {{Q\hspace{-0.15ex}U\/}}
\opname{U} {{U\/}}

\opname{cGQ} {{\cal G \hspace{-0.2em} Q \/}}
\opname{cSL} {{\cal S \hspace{-0.2em} L \/}}
\opname{cGL} {{\cal G \hspace{-0.2em} L \/}}
\opname{cOSp} {{\cal O \hspace{-0.2em} S \hspace{-0.3em} \it p\/}}
\opname{cPe} {{\cal P \hspace{-1.5pt} \it e\/}}
\opname{cVect} {{\cal V \hspace{-1.5pt} \it e\hspace{-0.1ex}c\hspace{-0.1ex}t\/}}
\opname{cVol} {{\cal V \hspace{-1.5pt} \it o\hspace{-0.1ex}l\/}}
\opname{cAut} {{\cal A \hspace{-0.2em} \it u\hspace{-0.1em}t\/}}
\opname{cCovect} {{\cal C \hspace{-1.5pt}
     \it o\hspace{-0.1ex}v\hspace{-0.1ex}e\hspace{-0.1ex}c\hspace{-0.1ex}t\/}}
\opname{CW} {{C\hspace{-0.15ex}W}}

\opname {Ab}   {{\sf Ab}}
\opname {Alg}   {{\sf Alg}}
\opname {ASch}  {{\sf Aff\;Sch}}
\opname {Funct}   {{\sf Funct}}
\opname {Gr}   {{\sf Gr}}
\opname {Grf}  {{{\sf Gr}_f}}
\opname {Mods}   {{\sf mods}}
\opname {Rings}   {{\sf Rings}}
\opname {Salg}   {{\sf Salg}}
\opname {Sets} {{\sf Sets}}
\opname {SSMan} {{\sf SMan}}
\opname {Top}  {{\sf Top}}
\opname {Vebun}   {{\sf Vebun}}

\newcommand {\ev} {{\bar0}}
\newcommand {\od} {{\bar1}}

\newcommand {\degree}  {{}^\circ}
\newcommand {\tto} {\longrightarrow}
\newcommand {\pder}[1] {{\frac{\partial}{\partial {#1}}}}
\newcommand {\pderf}[2] {{\frac{\partial {#1}}{\partial {#2}}}}

\newcommand {\bcdot}   {\mathbin{\hbox{\raise.4ex\hbox{\bf.}}}} 

\newcommand {\secno} {}
\newcommand {\ssecfont} {\normalfont\bf}

\newtheorem{Theorem}{\secno Theorem}

\newtheorem{Lemma}[Theorem]{\secno Lemma}

\newtheorem{Corollary}[Theorem]{\secno Corollary}

\newenvironment {th*}[1]
    {\gdef\thname{#1} \begin{thn}}%
    {\end{thn}}
\newtheorem{thn}[Theorem] {\thname}

\theoremstyle{definition}

\newtheorem{Convention}[Theorem]{\secno Convention}

\newenvironment {ex*}[1]
    {\gdef\thname{#1} \begin{exn}}%
    {\end{exn}}
\newtheorem{exn}[Theorem]{\thname}

\theoremstyle{remark}

\newenvironment {rem*}[1]
    {\gdef\thname{#1} \begin{remn}}%
    {\end{remn}}
\newtheorem{remn}[Theorem]{\thname}

\newcommand {\ssec}{\subsection*}

\newcommand {\ssbegin}[2]
  {\def \secno {\gdef \secno {}{\ssecfont #1. }}%
   \begin{#2}}
\begin{document}

\title{The five exceptional simple Lie superalgebras of vector fields}

\author{Irina Shchepochkina} 

\address{On leave of absence from the Independent Univ. of Moscow.
Correspondence: c/o D. Leites, Dept. of Math., Univ. of Stockholm, Roslagsv.
101, Kr\"aftriket hus 6, S-106 91, Stockholm, Sweden}


\begin{abstract} The five simple exceptional complex Lie superalgbras of
vector fields are described. One of them is new; the other four are explicitely
described for the first time. All of the exceptional Lie superalgebras are obtained with
the help of the Cartan prolongation or a generalized prolongation. 

The description of several of the exceptional Lie superalgebras is
associated with the Lie superalgebra $\fas$ --- the nontrivial central
extension of the supertraceless subalgebra $\fspe(4)$ of the {\it
periplectic} Lie superalgebra $\fpe(4)$ that preserves the nondegenerate odd
bilinear form on the $4|4$-dimensional superspace. (A nontrivial central
extension of $\fspe(n)$ only exists for $n=4$.) 
\end{abstract} 

\thanks{I am thankful to D.~Leites for rising the problem and help; to RFBR grant 
95-01187 01 and NFR for financial support; University of Twente and Stockholm
University for hospitality. Computer experiments by G.~Post and P.~Grozman encouraged
me to carry on with unbearable calculations.}

\maketitle

\section*{Introduction}

V. Kac [K] classified simple finite dimensional Lie superalgebras
over ${\Cee}$. Kac further conjectured [K] that passing to infinite dimensional
simple Lie superalgebras of vector fields with polynomial coefficients we only
acquire the straightforward analogs of the four well-known Cartan
series $\fvect(n)$, $\fs\fvect(n)$, $\fh(2n)$ and $\fk(2n+1)$ (of all,
divergence-free, hamiltonian and contact vector fields, respectively, realized on the
space of dimension indicated). Since superdimension is a pair of numbers, Kac's 
examples of simple vectoral Lie superalgebras double Cartan's list of simple vectoral
Lie algebras.

It soon became clear [L1], [ALSh] that the actual list of simple vectoral
Lie superalgebras \lq\lq doubles" that of Cartan twice, not once (the super counterparts
$\fm$ and $\fsm$ of $\fk$ as well as $\fle$ and $\fsle$ --- the super counterparts of $\fh$
--- were discovered); moreover, even the superalgebras of the 4 well-known series
($\fvect$, $\fs\fvect$, $\fh$ and $\fk$) have, in addition to the dimension of the
superspace on which they are usually realized, one more discrete parameter indicating
other, nonstandard realizations. Furthermore, several of these Lie superalgebras have
deformations, see [L2], [L3]. 

Next, three exceptional vectoral algebras were discovered [Sh1], followed by
a fourth exception [Sh2]. The purpose of this note is to give a more lucid
description of these exceptions, and introduce the most remarkable 5th
exception. (For a related construction of Lie superalgebras of string theories
see [GLS]; for further study of some of these superalgebras see [LSX] and [LS].)

In this note the ground field is $\Cee$. First, we recall the background
from Linear Algebra in Superspaces. Then we recall the definition of the main
tool in the construction of our examples: the notion of Cartan prolongation and
its generalization ([Sh1]). We recall the known facts form the classification of
simple Lie superalgebras of vector fields, cf. [L2] and [L3]. 

{\bf Our main result} is the discovery and a description of the five exceptional simple
Lie algebras of vector fields: $(\spin, \fas)_*$, $(\Pi (T(\vec 0)/{\Cee}\cdot
1), \fcvect (0|3))_*$, $(\fab(4), \fcvect (0|3))_*^{m}$,  $(\fhei(8|6),
\fsvect _{3, 4}(4))_*^{k}$ and $\fk\fas$. These names reflect the method of construction of
these algebras, rather than their own properties; for private use I call them shortly $\fx_1$
-- $\fx_5$, respectively. To name them properly, an interpretation is required.

The notions of the generalized and a {\it partial} Cartan prolongations are the key tools
in the construction.

I divide the description into several sections, according to the method
of construction. Boring calculations are gathered in Appendix. The statements on simplicity are
proved via Kac's criteria, cf. [K]. Proof of the fact that the five exceptional Lie
superalgebras have only one realization and the clasification of their real forms will be
given elsewhere.

The reader might wonder if new simple Lie superalgebras of vector fields will continue
to appear every several years. We conjecture that there is no such
treat/danger any longer: the classification of graded Lie superalgebras of polynomial
growth and finite depth is soon to appear, at last [LSh].

{\bf Open problems}: (1) Give geometric realizations of the
exceptional Lie superalgebras (what structures do they preserve?). 

(2) Certain
exceptional Lie superalgebra are deformations of (nonsimple) Lie superalgebras
whose brackets are easy to describe. In this paper the cocycle is described clumsily,
in components of the generating functions. Describe the cocycle (i.e.,
the bracket itself) in terms of generating functions themselves. (An attempt is made in
[PSh].)

(3) Find out what do our
exceptional Lie superalgebras add to the list of simple finite dimensional Lie
algebras over an algebraically closed field of characteristic 2 via {\it Leites'
conjecture} [KL], [L2].

\vskip 0.2 cm
{\bf Remark}. The results of this paper and the related results on classification of
the stringy superalgebras [GLS] were obtained in Stockholm in June 1996 and delivered at
the seminar of E.~Ivanov, JINR, Dubna (July, 1996) and Voronezh winter school Jan. 12--18,
1997. The interest of V.~Kac during October--November 1996 stimulated me to finish
editing this paper. I am particularly thankful to P.~Grozman, D.~Leites and G.~Post who
helped me.

\section*{\protect \S 0. Background}

\ssec{0.1. Linear algebra in superspaces. Generalities}
Superization has certain subtleties, often disregarded or expressed as in [L],
[L3] or [M]: too briefly. We will dwell on them a bit.

A {\it superspace} is a $\Zee /2$-graded space; for a superspace
$V=V_{\ev}\oplus V_{\od }$ denote by $\Pi (V)$ another copy of the same
superspace: with the shifted parity, i.e., $(\Pi(V))_{\bar i}= V_{\bar i+\od
}$. The {\it superdimension} of $V$ is $\dim V=p+q\varepsilon$, where
$\varepsilon^2=1$ and $p=\dim V_{\ev}$,
$q=\dim V_{\od }$. (Usually $\dim V$ is expressed as a pair $(p, q)$ or
$p|q$; this obscures the fact that $\dim V\otimes W=\dim V\cdot \dim W$ which is clear
with the use of $\varepsilon$.)

A superspace structure in $V$ induces the superspace structure in the space $\End
(V)$. A {\it basis of a superspace} is always a basis consisting of {\it
homogeneous} vectors; let
$\Par=(p_1, \dots, p_{\dim V})$ be an ordered collection of their parities.
We call $\Par$ the {\it format} of the basis of $V$. A square {\it supermatrix} of format
(size)
$\Par$ is a $\dim V\times \dim V$ matrix whose $i$th row and $i$th column are of
the same parity $p_i$. The matrix unit $E_{ij}$ is supposed to be of parity
$p_i+p_j$ and the bracket of supermatrices (of the same format) is defined via Sign Rule: {\it if something of
parity $p$ moves past something of parity $q$ the sign $(-1)^{pq}$ accrues; the
formulas defined on homogeneous elements are extended to arbitrary ones via
linearity}. For example: setting $[X, Y]=XY-(-1)^{p(X)p(Y)}YX$ we get the notion
of the supercommutator and the ensuing notion of the Lie superalgebra (that
satisfies the superskew-commutativity and super Jacobi identity).

We do not usually use the sign $\wedge$ for the wedge product of differential forms on
supermanifolds: in what follows we assume that the exterior differential is odd and the
differential forms constitute a supercommutative superalgebra; we keep using it on
manifolds, sometimes, not to diviate too far from conventional notations. 

Usually, $\Par$ is of the form $(\ev , \dots, \ev , \od , \dots, \od )$. Such a
format is called {\it standard}. In this paper we can do without nonstandard formats. But
they are vital in the study of systems of simple roots that the reader might be
interested in.

The {\it general linear} Lie superalgebra of all supermatrices of size $\Par$ is
denoted by $\fgl(\Par)$; usually, $\fgl(\ev, \dots, \ev, \od, \dots, \od)$ is
abbreviated to $\fgl(\dim V_{\bar 0}|\dim V_{\bar 1})$. Any matrix from
$\fgl(\Par)$ can be expressed as the sum of its even
and odd parts; in the standard format this is the block expression: 
$$
\begin{pmatrix}A&B\\ C&D\end{pmatrix}=\begin{pmatrix}A&0\\
0&D\end{pmatrix}+\begin{pmatrix}0&B\\ C&0\end{pmatrix},\quad p\left(\begin{pmatrix}A&0\\
0&D\end{pmatrix}\right)=\ev, \; p\left(\begin{pmatrix}0&B\\
C&0\end{pmatrix}\right)=\od.
$$

The {\it supertrace} is the map $\fgl (\Par)\longrightarrow \Cee$,
$(A_{ij})\mapsto \sum (-1)^{p_{i}}A_{ii}$. Since $\str [x, y]=0$, the space of
supertraceless matrices constitutes the {\it special linear} Lie subsuperalgebra
$\fsl(\Par)$.

{\bf Superalgebras that preserve bilinear forms: two types}. To the linear map $F$ of 
superspaces there corresponds the dual map $F^*$ between the
dual superspaces; if $A$ is the supermatrix corresponding to $F$ in a basis of the format
$\Par$, then to $F^*$ the {\it supertransposed} matrix $A^{st}$ corresponds:
$$
(A^{st})_{ij}=(-1)^{(p_{i}+p_{j})(p_{i}+p(A))}A_{ji}.
$$

The supermatrices $X\in\fgl(\Par)$ such that 
$$
X^{st}B+(-1)^{p(X)p(B)}BX=0\quad \text{for an homogeneous matrix $B\in\fgl(\Par)$}
$$
constitute the Lie superalgebra $\faut (B)$ that preserves the bilinear form on $V$ with
matrix $B$. Most popular is the nondegenerate supersymmetric form whose matrix in the
standard format is the canonical form $B_{ev}$ or $B'_{ev}$:
$$
B_{ev}(m|2n)= \begin{pmatrix} 
1_m&0\\
0&J_{2n}
\end{pmatrix},\quad \text{where $J_{2n}=\begin{pmatrix}0&1_n\\-1_n&0\end{pmatrix}$,
or}\; B'_{ev}(m|2n)= \begin{pmatrix} 
\antidiag (1, \dots , 1)&0\\
0&J_{2n}
\end{pmatrix}. 
$$
The usual notation for $\faut (B_{ev}(m|2n))$ is $\fosp(m|2n)$ or $\fosp^{sy}(m|2n)$.
(Observe that the passage from $V$ to $\Pi (V)$ sends the supersymmetric forms to
superskew-symmetric ones, preserved by the \lq\lq
symplectico-orthogonal" Lie superalgebra $\fsp'\fo (2n|m)$ or $\fosp^{sk}(m|2n)$ which
is isomorphic to
$\fosp^{sy}(m|2n)$ but has a different matrix realization. We never use notation
$\fsp'\fo (2n|m)$ in order not to confuse with the special Poisson
superalgebra.)

In the standard format the matrix realizations of these algebras
are: 
$$
\begin{matrix} 
\fosp (m|2n)=\left\{\left (\begin{matrix} E&Y&-X^t\\
X&A&B\\
Y^t&C&-A^t\end{matrix} \right)\right\};\quad \fosp^{sk}(m|2n)=
\left\{\left(\begin{matrix} A&B&X\\
C&-A^t&Y^t\\
Y^t&-X^t&E\end{matrix} \right)\right\}, \\
\text{where}\; 
\left(\begin{matrix} A&B\\
C&-A^t\end{matrix} \right)\in \fsp(2n),\qquad E\in\fo(m)\;
\text{and}\;  {}^t \; \text{is the usual transposition}.\end{matrix} 
$$

A nondegenerate supersymmetric odd bilinear form $B_{odd}(n|n)$ can be
reduced to the canonical form whose matrix in the standard format is 
$J_{2n}$. A canonical form of the superskew odd nondegenerate form in the
standard format is $\Pi_{2n}=\begin{pmatrix} 0&1_n\\1_n&0\end{pmatrix}$.
The usual notation for $\faut (B_{odd}(\Par))$ is $\fpe(\Par)$. The passage
from $V$ to $\Pi (V)$ sends the supersymmetric forms to superskew-symmetric
ones and establishes an isomorphism $\fpe^{sy}(\Par)\cong\fpe^{sk}(\Par)$. This Lie
superalgebra is called, as A.~Weil suggested, {\it periplectic}, i.e., odd-plectic. The
In the standard format these superalgebras are shorthanded as
following formula, where their matrix realizations is also given:
$$
\begin{matrix}
\fpe ^{sy} (n)=\left\{\begin{pmatrix} A&B\\
C&-A^t\end{pmatrix}, \; \text{where}\; B=-B^t,
C=C^t\right\};\\
\fpe^{sk}(n)=\left\{\begin{pmatrix}A&B\\ C&-A^t\end{pmatrix}, \;
\text{where}\; B=B^t, C=-C^t\right\}.
\end{matrix}
$$

The {\it special periplectic} superalgebra is $\fspe(n)=\{X\in\fpe(n): \str
X=0\}$.

Observe that though the Lie superalgebras $\fosp^{sy} (m|2n)$ and $\fpe ^{sk} (2n|m)$, as
well as $\fpe ^{sy} (n)$ and $\fpe ^{sk} (n)$, are isomorphic, the difference between them
is sometimes crucial, see Remark 0.6 below.

\ssec{0.2. Vectoral Lie superalgebras. The standard realization} The elements of $\cL=\fder\;
\Cee [[u]]$ are considered as vector fields. The Lie algebra $\cL$ has only one maximal
subalgebra
$\cL_0$ of finite codimension (consisting of the fields that vanish at the origin). The
subalgebra $\cL_0$ determines a filtration of $\cL$: set
$$
\cL_{-1}=\cL;\quad \cL_i =\{D\in \cL_{i-1}: [D, \cL]\subset\cL_{i-1}\}\; \text{for
}i\geq 1.
$$
The associated graded Lie algebra $L=\mathop{\oplus}\limits_{i\geq -1}L_i$, where
$L_i=\cL_{i}/\cL_{i+1}$, consists of the vector fields with {\it polynomial} coefficients. 

Suppose $\cL_0 subset\cL$ is a maximal subalgebra of finite codimension and containing no
ideals of $\cL$. For the Lie superalgebra $\cL=\fder\, \Cee [u, \xi]$ the minimal subspace of
$\cL$ containing $\cL_0$ coincides with $\cL$. Not all the subalgebras $\cL$ of $\fder\, \Cee
[u, \xi]$ have this property. Let $\cL_{-1}$ be a minimal subspace of
$\cL$ containing $\cL_0$, different from $\cL_0$ and $\cL_0$-invariant. Construct a filtration
of $\cL$ by setting
$$
\cL_{-i-1}=[\cL_{-1},
\cL_{-i}]+\cL_{-i}\;\quad \cL_i =\{D\in \cL_{i-1}: [D, \cL{-1}]\subset\cL_{i-1}\}\; 
\text{for }i>0.
$$
Since the codimension of $\cL_0$ is finite, the filtration takes the form
$$
\cL=\cL_{-d}\supset\dots\cL_{0}\supset\dots \eqno{(*)}
$$
for some $d$. This $d$ is the {\it depth} of $\cL$ or the associated graded Lie superalgebra.
We call all filtered or graded Lie superalgebras of finite depth {\it vectoral}, i.e.,
realizable with vector fields on a finite dimensional supermanifold. Considering the
subspaces $(*)$ as the basis of a topology, we can complete the graded or filtered Lie
superalgebras $L$ or $\cL$; the elements of the completion are the vector fields with formal
power series as coefficients. Though the structure of the graded algebras is easier to
describe, in applications the completed Lie superalgebras are usually needed. 

Unlike Lie algebras, simple vectoral {\it super}algebras possess {\it several} maximal
subalgebras of finite codimension. We describe them, together with the corresponding
gradings, in subsect. 0.4. 

{\bf 1) General algebras}. Let $x=(u_1, \dots , u_n, \theta_1, \dots ,
\theta_m)$, where the $u_i$ are even indeterminates and the $\theta_j$ are odd ones.
The Lie superalgebra $\fvect (n|m)$ is $\fder\; \Cee[x]$; it is called {\it the
general vectoral superalgebra}. \index{$\fvect$ general vectoral Lie
superalgebra}\index{ Lie superalgebra general vectoral}

{\bf 2) Special algebras}. The {\it divergence}\index{divergence} of the field
$D=\sum\limits_if_i\pder{u_{i}} + \sum\limits_j
g_j\pder{\theta_{j}}$ is the function (in our case: a
polynomial, or a series) 
$$
\Div D=\sum\limits_i\pderf{f_{i}}{u_{i}}+
\sum\limits_j (-1)^{p(g_{j})}
\pderf{g_{i}}{\theta_{j}}.
$$

$\bullet$ The Lie superalgebra $\fsvect (n|m)=\{D \in \fvect (n|m): \Div D=0\}$
is called the {\it special} or {\it divergence-free vectoral superalgebra}.
\index{$\fsvect$ general vectoral Lie superalgebra}\index{ Lie superalgebra
special vectoral}\index{ Lie superalgebra divergence-free} It is not difficult to see that
it is also possible to describe $\fsvect$ as $\{ D\in \fvect (n|m): L_D\vvol _x=0\}$, where
$\vvol_x$ is the volume form with constant coefficients in coordinates $x$ and $L_D$ the
Lie derivative with respect to
$D$. 

$\bullet$ The Lie superalgebra $\fsvect_{\lambda}(0|m)=\{D \in \fvect (0|m):
\Div (1+\lambda\theta_1\cdot \dots \cdot
\theta_m)D=0\}$ --- the deform of $\fsvect(0|m)$ --- is called the
{\it special} or {\it divergence-free vectoral superalgebra}. Clearly, that 
$\fsvect_{\lambda}(0|m)\cong \fsvect_{\mu}(0|m)$ for
$\lambda\mu\neq 0$. Observe that $p(\lambda)\equiv m\pmod 2$, i.e., for odd $m$
the parameter of deformation $\lambda$ is odd.

\begin{rem*}{Remark} Sometimes we write $\fvect (x)$ or even $\fvect (V)$ if
$V=\Span(x)$ and use similar notations for the subalgebras of $\fvect$ introduced
below. Algebraists sometimes abbreviate $\fvect (n)$ and $\fs\fvect (n)$ to $W_n$
(in honor of Witt) and $S_n$, respectively.
\end{rem*}

{\bf 3) The algebras that preserve Pfaff
equations and differential 2-forms}. 

$\bullet$ Set $u=(t, p_1, \dots , p_n, q_1, \dots , q_n)$; let
$$
\tilde \alpha_1 = dt +\sum\limits_{1\leq i\leq n}(p_idq_i - q_idp_i)\ +
\sum\limits_{1\leq j\leq m}\theta_jd\theta_j\quad\text{and}\quad 
 \omega_0=d\alpha_1\ .
$$
The form $\tilde \alpha_1$ is called {\it contact}, the form  $\tilde \omega_0$ is called {\it
symplectic}.\index{form differential contact}\index{form differential
symplectic} Sometimes it is more convenient to redenote
the $\theta$'s and set 
$$
\xi_j=\frac{1}{\sqrt{2}}(\theta_{j}-i\theta_{r+j});\quad \eta_j=\frac{1}{
\sqrt{2}}(\theta_{j}+i\theta_{r+j})\; \text{ for}\; j\leq r= [m/2]\; (\text{here}\;
i^2=-1),
\quad
\theta =\theta_{2r+1} 
$$ 
and in place of $\tilde \omega_0$ or $\tilde \alpha_1$ take $\alpha$
and $\omega_0=d\alpha_1$, respectively, where 
$$
\begin{array}{rcl}
\alpha_1=&dt+\sum\limits_{1\leq i\leq n}(p_idq_i-q_idp_i)+
\sum\limits_{1\leq j\leq r}(\xi_jd\eta_j+\eta_jd\xi_j)&
\text{ if }\ m=2r\\
\alpha
_1=&dt+\sum\limits_{1\leq i\leq n}(p_idq_i-q_idp_i)+
\sum\limits_{1\leq j\leq r}(\xi_jd\eta_j+\eta_jd\xi_j) +\theta d\theta&\text{ if
}\ m=2r+1.\end{array} 
$$

The Lie superalgebra that preserves the {\it Pfaff equation}
\index{Pfaff equation} $\alpha_1=0$, i.e., the superalgebra
$$
\fk (2n+1|m)=\{ D\in \fvect (2n+1|m): L_D\alpha_1=f_D\alpha_1\}, 
$$
(here $f_D\in \Cee [t, p, q, \xi]$ is a polynomial determined by $D$) is
called the {\it contact superalgebra}.\index{$\fk$ contact superalgebra}
\index{Lie superalgebra contact} The Lie superalgebra that preserves not just the Pfaff
equation determined by $\alpha_1$ but the form itself, i.e.,
$$
\begin{array}{c}
\fpo (2n|m)=\{ D\in \fk (2n+1|m): L_D\alpha_1=0\}\end{array}
$$
is called the {\it Poisson} superalgebra.\index{$\fpo$ Poisson superalgebra}
(A geometric interpretation of the
Poisson superalgebra: it is the Lie superalgebra that preserves the connection
with form $\alpha_1$ in the line bundle over a symplectic supermanifold
with the symplectic form $d\alpha_1$.) 

$\bullet$ Similarly, set $u=q=(q_1, \dots , q_n)$,
let $\theta=(\xi_1, \dots , \xi_n; \tau)$ be odd. Set
$$
\begin{array}{c}
\alpha_0=d\tau+\sum\limits_i(\xi_idq_i+q_id\xi_i), \qquad\qquad
\omega_1=d\alpha_0
\end{array}
$$
and call these forms the {\it odd contact} and {\it periplectic}, 
respectively.\index{form differential contact odd}\index{form differential
periplectic}

The Lie superalgebra that preserves
the Pfaff equation $\alpha_0=0$, i.e., the superalgebra 
$$
\fm (n)=\{ D\in \fvect (n|n+1):L_D\alpha_0=f_D\cdot \alpha_0\}, \; \text{
where }\; f_D\in \Cee [q, \xi, \tau], 
$$
is called the {\it odd contact superalgebra}.\index{$\fm$ contact
superalgebra}

The Lie superalgebra \index{$\fb$ Buttin superalgebra}
$$
\begin{array}{c}
\fb (n)=\{ D\in \fm (n): L_D\alpha_0=0\}\end{array}
$$
is called the {\it Buttin} superalgebra ([L3]). (A geometric interpretation of the
Buttin superalgebra: it is the Lie superalgebra that preserves
the connection
with form $\alpha_1$ in the line bundle of rank $\varepsilon$ over a
periplectic supermanifold, i.e., the supermanifold with the periplectic
form $d\alpha_0$.)

The Lie superalgebras
$$
\begin{array}{c}
\fsm (n)=\{ D\in \fm (n): \Div\ D=0\}\ , \ \fsb (n)=\{ D\in
\fb (n):\Div\ D=0\} 
\end{array}
$$
are called the {\it divergence-free} (or {\it special}) {\it odd contact} and
{\it special Buttin} superalgebras, respectively.

\begin{rem*}{Remark} A relation with finite dimensional geometry is as follows.
Clearly, $\ker \alpha_1= \ker \alpha _1$. The restriction of $\omega_0$ to
$\ker 
\alpha_1$ is the orthosymplectic form $B_{ev}(m|2n)$; the restriction of
$\omega_0$ to $\ker \alpha_1 $ is $B'_{ev}(m|2n)$. Similarly, the restriction of
$\omega_1$ to $\ker \alpha_0$ is the periplectic form $B_{odd}(n|n)$.
\end{rem*}

\ssec{0.3. Generating functions} A laconic way to describe the elements of
$\fk$, $\fm$ and their subalgebras is via generating functions.

$\bullet$ Odd form $\alpha_1$. For $f\in\Cee [t, p, q, \xi]$ set\index{$K_f$
contact vector field} \index{$H_f$ Hamiltonian vector field}: 
$$
K_f=\triangle(f)\pder{t}-H_f +
\pderf{f}{t} E, 
$$
where
$E=\sum\limits_i y_i
\pder{y_{i}}$ (here the $y$ are all the
coordinates except $t$) is the {\it Euler operator} (which counts the degree
with respect to the $y$), $\triangle (f)=2f-E(f)$, and $H_f$ is the
hamiltonian field with Hamiltonian $f$ that preserves $d\alpha_1$: 
$$
H_f=\sum\limits_{i\leq n}(\pderf{f}{p_i}
\pder{q_i}-\pderf{f}{q_i}
\pder{p_i}) -(-1)^{p(f)}\left(\sum\limits_{j\leq m}\pderf{
f}{\theta_j} \pder{\theta_j}\right ) , \; \; f\in \Cee [p,
q, \theta]. 
$$

The choice of the form $\alpha_1$ instead of $\tilde\alpha_1$ only affect the
form of $H_f$ that we give for $m=2k+1$:
$$
H_f=\sum\limits_{i\leq n} (\pderf{f}{p_i}
\pder{q_i}-\pderf{f}{q_i}
\pder{p_i}) -(-1)^{p(f)}\sum\limits_{j\leq
k}(\pderf{f}{\xi_j} \pder{\eta_j}+
\pderf{f}{\eta_j} \pder{\xi_j}+
\pderf{f}{\theta} \pder{\theta}), \;
\; f\in \Cee [p, q, \xi, \eta, \theta]. 
$$

$\bullet$ Even form $\alpha_0$. For $f\in\Cee [q, \xi, \tau]$ set:
$$
M_f=\triangle(f)\pder{\tau}- Le_f
-(-1)^{p(f)} \pderf{f}{\tau} E, 
$$
where $E=\sum\limits_iy_i 
\pder{y_i}$ (here the $y$ are all the coordinates except
$\tau$) is the Euler operator, $\triangle(f)=2f-E(f)$, and
$$
Le_f=\sum\limits_{i\leq n}( \pderf{f}{q_i}\ 
\pder{\xi_i}+(-1)^{p(f)} \pderf{f}{\xi_i}\ 
\pder{q_i}), \; f\in \Cee [q, \xi].
$$
\index{$M_f$ contact vector field} \index{$Le_f$ periplectic vector field} 

Since 
$$
L_{K_f}(\alpha_1)=2 \pderf{f}{t}\alpha_1, \quad\quad
L_{M_f}(\alpha_0)=-(-1)^{p(f)}2 \pderf{
f}{\tau}\alpha_0,
$$
it follows that $K_f\in \fk (2n+1|m)$ and $M_f\in \fm (n)$. Observe that
$$
p(Le_f)=p(M_f)=p(f)+\od.
$$

$\bullet$ To the (super)commutators $[K_f, K_g]$ or $[M_f, M_g]$ there
correspond {\it contact brackets}\index{Poisson bracket}\index{contact bracket}
of the generating functions:
$$
[K_f, K_g]=K_{\{f, g\}_{k.b.}};\quad\quad [M_f, M_g]=M_{\{f, g\}_{m.b.}}.
$$
The explicit formulas for the contact brackets are as follows. Let us first
define the brackets on functions that do not depend on $t$ (resp. $\tau$).

The {\it Poisson bracket} $\{\cdot , \cdot\}_{P.b.}$ (in the realization with the form
$\omega_0$) is given by the formula 
$$
\begin{array}{c}
\{f, g\}_{P.b.}=\sum\limits_{i\leq n}\ (\pderf{f}{p_i}\ 
\pderf{g}{q_i}-\ \pderf{f}{q_i}\ 
\pderf{g}{p_i})-(-1)^{p(f)}\sum\limits_{j\leq m}\ 
\pderf{f}{\theta_j}\ \pderf{g}{\theta_j}\end{array}
$$
and in the realization with the form
$\omega_0$ for $m=2k+1$ it is given by the formula 
$$
\begin{array}{c}
\{f, g\}_{P.b.}=\sum\limits_{i\leq n}\ (\pderf{f}{p_i}\ 
\pderf{g}{q_i}-\ \pderf{f}{q_i}\ 
\pderf{g}{p_i})-(-1)^{p(f)}[\sum\limits_{j\leq m}( 
\pderf{f}{\xi_j}\ \pderf{
g}{\eta_j}+\pderf{f}{\eta_j}\ \pderf{
g}{\xi_j})+\pderf{f}{\theta}\ \pderf{
g}{\theta}].
\end{array} 
$$

The {\it Buttin bracket} $\{\cdot ,
\cdot\}_{B.b.}$ \index{Buttin bracket $=$ Schouten bracket} is given by the formula
$$
\{ f, g\}_{B.b.}=\sum\limits_{i\leq n}\ (\pderf{f}{q_i}\ 
\pderf{g}{\xi_i}+(-1)^{p(f)}\ \pderf{f}{\xi_i}\ 
\pderf{g}{q_i}).
$$

\footnotesize
\begin{rem*}{Remark} The what we call here Buttin bracket was
discovered in pre-super era by Schouten. Buttin was the
first to observe that the Schouten bracket determines a Lie superalgebra. The {\it Schouten
bracket} was originally defined on the superspace of polyvector fields on a manifold, i.e.,
on the superspace $\Gamma(\Lambda^{\bcdot}(T(M)))\cong\Lambda^{\bcdot}_\cF(Vect(M))$ of
sections of the exterior algebra (over the algebra $\cF$ of functions) of the tangent
bundle. The explicit formula of the Schouten bracket (in which the hatted slot should be
ignored, as usual) is 
$$
[X_1\wedge\dots \wedge\dots \wedge X_k, Y_1\wedge\dots \wedge Y_l]=
\sum_{i, j}(-1)^{i+j}[X_i, Y_j]\wedge X_1\wedge\dots\wedge \hat X_i\wedge 
\dots\wedge X_k\wedge Y_1\wedge\dots\wedge \hat Y_j\wedge \dots \wedge
Y_l.\eqno{(*)}
$$
With the help of Sign Rule we easily superize formula $(*)$ for the case
when manifold $M$ is replaced with supermanifold $\cM$. Let $x$ and $\xi$
be the even and odd coordinates on $\cM$. Setting $\theta_i=\Pi(\partial
{x_{i}})=\check x_{i}$, $q_j=\Pi(\partial {\xi_{j}})=\check \xi_{j}$ we get an
identification of the Schouten bracket of polyvector fields on $\cM$ with the
Buttin bracket of functions on the supermanifold $\check\cM$; for the
definition of $\check\cM$, see [M]. (Physicists call checked variables {\it ghosts}
[GPS].)
\end{rem*}
\normalsize

In terms of the Poisson and Buttin brackets, respectively, the contact
brackets take the form
$$
\{ f, g\}_{k.b.}=\triangle (f)\pderf{g}{t}-\pderf{f}
{t}\triangle (g)-\{ f, g\}_{P.b.}
$$
and, respectively,
$$
\{ f, g\}_{m.b.}=\triangle (f)\pderf{g}{\tau}+(-1)^{p(f)}
\pderf{f}{\tau}\triangle (g)-\{ f, g\}_{B.b.}.
$$

The Lie superalgebras of {\it Hamiltonian fields}\index{Hamiltonian
vector fields} (or {\it Hamiltonian 
superalgebra}) and its special subalgebra (defined only if $n=0$) are
$$
\fh (2n|m)=\{ D\in \fvect (2n|m):\ L_D\omega_0=0\}\; \text{ and} \;
\fsh (m)=\{ D\in \fh (0|m): \Div D=0\}.
$$
Its odd analogues are the Lie superalgebra of Leitesian fields introduced in
[L1] and its special subalgebra:
$$
\fle (n)=\{ D\in \fvect (n|n): L_D\omega_1=0\} \; \text{ and} \;
\fsle (n)=\{ D\in \fle (n): \Div D=0\}.
$$

It is not difficult to prove the following isomorphisms (as superspaces): 
$$
\begin{array}{rclrcl}
\fk (2n+1|m)&\cong&\Span(K_f: f\in \Cee[t, p, q, \xi]);&\fle
(n)&\cong&\Span(Le_f: f\in \Cee [q, \xi]);\\
\fm (n)&\cong&\Span(M_f: f\in \Cee [\tau, q, \xi]);&
\fh (2n|m)&\cong&\Span(H_f: f\in
\Cee [p, q, \xi]).
\end{array}
$$

\begin{rem*}{Remark} 1) It is obvious that the Lie superalgebras of the
series $\fvect$, $\fsvect$, $\fh$ and $\fpo$ for $n=0$ are finite dimensional.

2) A Lie superalgebra of the series $\fh$ is the quotient of the Lie
superalgebra $\fpo$ modulo the one-dimensional center
$\fz$ generated by constant functions.
Similarly, $\fle$ and $\fsle$ are the quotients of $\fb$ and $\fsb$,
respectively, modulo the one-dimensional (odd) center $\fz$ generated by
constant functions. 
\end{rem*}

Set $\fspo (m)=\{ K_f\in \fpo (0|m):\int fv_\xi=0\}$; clearly, $\fsh (m)=\fspo
(m)/\fz$.

Since
$$
\Div M_f =(-1)^{p(f)}2\left ((1-E)\pderf{f}{\tau} - \sum\limits_{i\leq
n}\frac{\partial^2 f}{\partial q_i \partial\xi_i}\right ), 
$$
it follows that
$$
\fsm (n) = \Span\left (M_f \in \fm (n): (1-E)\pderf{f}{\tau}
=\sum\limits_{i\leq n}\frac{\partial^2 f}{\partial q_i
\partial\xi_i}\right ).
$$

In particular, 
$$
\Div Le_f = (-1)^{p(f)}2\sum\limits_{i\leq n}\frac{\partial^2 f}{\partial 
q_i \partial\xi_i}.
$$
The odd analog of the Laplacian, namely, the operator
$$
\Delta=\sum\limits_{i\leq n}\frac{\partial^2 }{\partial 
q_i \partial\xi_i}
$$
on a periplectic supermanifold appeared in physics under the name of {\it
BRST operator}, cf. [GPS]. The divergence-free vector fields from $\fsle(n)$ are generated
by {\it harmonic} functions, i.e., such that $\Delta(f)=0$.

Lie superalgebras $\fsle (n)$, $\fs\fb (n)$ and $\fsvect (1|n)$
have ideals $\fsle \degree(n)$, $\fs\fb \degree(n)$ and $\fsvect
\degree(n)$ of codimension 1 defined from the exact sequences 
$$ 
\begin{matrix}
0\longrightarrow \fsle \degree(n)\longrightarrow \fsle (n)\longrightarrow \Cee\cdot
Le_{\xi_1\dots\xi_n} \longrightarrow 0, \\
0\longrightarrow \fs\fb \degree(n)\longrightarrow \fs\fb (n)\longrightarrow \Cee\cdot
M_{\xi_1\dots\xi_n} \longrightarrow 0, \\
0\longrightarrow \fsvect \degree(n)\longrightarrow \fsvect (1|n)\longrightarrow 
\Cee \cdot\xi_1\dots\xi_n\pder{t}\longrightarrow 0.\end{matrix}
$$

\ssec{0.4. Nonstandard realizations} 
In [LSh] we proved that  the following are all the nonstandard gradings of the Lie
superalgebras indicated. Moreover, the gradings in the series
$\fvect$ induce the gradings in the series $\fsvect$, and $\fsvect\degree$; the
gradings in $\fm$ induce the gradings in $\fsm$, $\fle$, $\fsle$,
$\fsle\degree$,
$\fb$, $\fsb$, $\fsb\degree$; the gradings in $\fk$ induce the gradings in
$\fpo$, $\fh$. In what follows we consider $\fk (2n+1|m)$ as preserving the Pfaff
eq. $\alpha =0$, where $\alpha
=dt+\sum{i\leq
n}(p_idq_i-q_idp_i)+\sum_{j\leq r}(\xi_jd\eta_j+\eta_jd\xi_j)+\sum_{k\leq
m-2r}\theta_kd\theta_k$. 

The standard realizations are marked by $(*)$ and in this case indication to $r=0$ is
omitted; note that (bar several exceptions for small
$m, n$) it corresponds to the case of the minimal codimension of ${\cal L}_0$.   
\small
$$
\renewcommand{\arraystretch}{1.3}
\begin{tabular}{|c|c|}
\hline
Lie superalgebra & its $\Zee$-grading \\ 
\hline
$\fvect (n|m; r)$, & $\deg u_i=\deg \xi_j=1$  for any $i, j$\hskip 5.5 cm
$(*)$\\ 
\cline{2-2}
$ 0\leq r\leq m$ & $\deg \xi_j=0$ for $1\leq j\leq r;\
\deg u_i=\deg \xi_{r+s}=1$ for any $i, s$ \\ 
\hline
 & $\deg \tau=2$, $\deg q_i=\deg 
\xi_i=1$  for any $i$ \hskip 4 cm $(*)$\\ 
\cline{2-2}
$\fm(n; r),\; 0\leq r\leq n$& $\deg \tau=\deg q_i=1$, $\deg 
\xi_i=0$ for any $i$\\ 
\cline{2-2}
& $\deg \tau=\deg q_i=2$, $\deg \xi_i=0$ for $1\leq i\leq r
<n$;\\ 
& $\deg u_{r+j}=\deg \xi_{r+j}=1$ for any $j$\\ 
\hline
$\fk (2n+1|m; r)$, $0\leq r\leq [\frac{m}{2}]$ & $\deg t=2$, $\deg p_i=\deg q_i=
\deg \xi_j=\deg \eta_j=\deg \theta_k=1$ for any $i, j, k$\qquad $(*)$ \\ 
\cline{2-2}
& $\deg t=\deg \xi_i=2$, $\deg 
\eta_{i}=0$ for $1\leq i\leq r\leq [\frac{m}{2}]$; \\
&$\deg p_i=\deg q_i=\deg \theta_{j}=1$ for
$j\geq 1$ and all $i$\\ 
\hline
$\fk(1|2m; m)$ & $\deg t =\deg \xi_i=1$, $\deg 
\eta_{i}=0$ for $1\leq i\leq m$ \\ \hline
\end{tabular}
$$
\vskip 0.2 cm

\normalsize
Observe that $\fk(1|2; 2)\cong\fvect(1|1)$ and $\fm(1; 1)\cong\fvect(1|1)$.

{\bf The exceptional nonstandard
gradings}. Denote the indeterminates and their respective exceptional degrees as follows
(here
$\fk(1|2)$ is considered in the realization that preserves the Pfaff eq. $\alpha_1=0$): 
$$
\begin{tabular}{|r|c|c|l|}
\hline
$\fvect(1|1)$&$t,  \xi$& 2, 1&1, $-1$\\
$\fk (1|2)$&$t, \xi,  \eta$& $1,  2,  -1$&\\
$\fm(1)$&$\tau,  q,  \xi$&1,  2,  -1&\\
\hline
\end{tabular}
$$
Denote the nonstandard exceptional realizations by indicating the above degrees after a
semicolon. We
get the following isomorphisms:
$$
\begin{array}{cc} 
\fvect (1|1; 2,  1) \cong \fk (1|2);\; \; &\fk (1|2; 1,  2,  -1) \cong \fm (1);\\  
\fvect (1|1; 1,  -1) \cong \fm (1);\; \; & \fm (1; 1,  2,  -1) \cong \fk (1|2). 
\end{array}
$$

Observe that the Lie superalgebras corresponding to different values of $r$ are isomorphic as
abstract Lie superalgebras, but as filtered ones they are distinct.

\ssec{0.5. Cartan prolongs} 
We will repeatedly use Cartan's prolongation. So
let me recall the definition and generalize it somewhat. Let $\fg$ be a Lie
algebra, $V$ a $\fg$-module, $S^i$ the operator of the $i$-th symmetric
power. Set $\fg_{-1} = V$, $\fg_0 = \fg$ 
and define the $i$-th {\it Cartan prolong} for $i > 0$ as

\begin{multline*}
\fg_i = \{X\in \Hom(\fg_{-1}, \fg_{i-1}): X(v)(w,...) = X(w)(v,...)\;
\text{ for any }\; v, w\in \fg_{-1}\}\\
= (S^i(\fg_{-1})^*\otimes \fg_0)\cap (S^{i+1}(\fg_{-1})^*\otimes \fg_{-1}). 
\end{multline*}

The {\it Cartan prolong} (the result of 
Cartan's {\it prolongation}) of the pair $(V, \fg)$ is $(\fg_{-1},
\fg_{0})_* = \mathop{\oplus}\limits_{i\geq -1} \fg_i$. (In what follows ${\bcdot}$ in
superscript denotes, as is now customary, the collection of all degrees, while $*$ is
reserved for dualization; in the subscripts we retain the oldfashioned $*$ instead of
${\bcdot}$ to avoid too close a contact with the punctuation marks.)
	
Suppose that the $\fg_0$-module $\fg_{-1}$ is faithful. Then, clearly, 

\begin{multline*}
(\fg_{-1}, \fg_{0})_*\subset \fvect (n) = \fder~
\Cee[x_1,..., x_n],\; \text{ where }\; n = dim~ \fg_{-1}\; \text{ and }\\
\fg_i = \{D\in \fvect(n): \deg D=i, [D, X]\in\fg_{i-1}\text{ for any }
X\in\fg_{-1}\}. 
\end{multline*}

It is subject to an easy verification that the Lie algebra structure on
$\fvect (n)$ induces same on $(\fg_{-1}, \fg_{0})_*$. 

Of the four simple vectoral Lie algebras, three are Cartan prolongs:
$\fvect(n)=(\id, \fgl(n))_*$, $\fsvect(n)=(\id, \fsl(n))_*$ and
$\fh(2n)=(\id, \fsp(n))_*$. The fourth one --- $\fk(2n+1)$ --- is also the
prolong under a trifle more general construction described as follows.

\ssec{A generalization of the Cartan prolongation} 
Let $\fg_-=\mathop{\oplus}\limits_{-d\leq i\leq -1}\fg_i$ be a nilpotent $\Zee$-graded Lie
algebra and $\fg_0\subset \fder_0\fg$ a Lie subalgebra of the $\Zee$-grading-preserving
derivations. For $i >0$ define the $i$-th prolong of the pair $(\fg_, \fg_0)$
to be: 
$$ 
\fg_i = ((S^{\bcdot}(\fg_-)^*\otimes \fg_0)\cap (S^{\bcdot}(\fg_-)^*\otimes
\fg_-))_i, 
$$ 
where the subscript $i$ in the rhs singles out the component of degree $i$. 

Define $\fg_*$, or rather, $(\fg_-, \fg_0)_*$, to be
$\mathop{\oplus}\limits_{i\geq -d}
\fg_i$; then, as is easy to verify, $(\fg_-, \fg_0)_*$ is a Lie algebra. 

What is the Lie algebra of contact vector fields in these terms? Denote by
$\fhei(2n)$ the Heisenberg Lie algebra: its space is $W\oplus {\Cee}\cdot z$,
where $W$ is a $2n$ dimensional space endowed with a nondegenerate skew-symmetric
bilinear form $B$ and the bracket in $\fhei(2n)$ is given by the
following conditions: $z$ is in the center and $[v, w]=B(v, w)\cdot z$ for any $v, w\in
W$.

Clearly, $\fk(2n+1)$ is
$(\fhei(2n), \fc\fsp(2n))_*$, where for any $\fg$ we write
$\fcg=\fg \oplus {\Cee}\cdot z$ or $\fc(\fg)$ to denote the trivial central
extension with the 1-dimensional even center generated by $z$.

\ssec{0.6. Lie superalgebras of vector fields as the Cartan prolongs} The
superization of the constructions from sec. 0.5 are straightforward: via Sign
Rule. We thus get:
$\fvect(m|n)=(\id, \fgl(m|n))_*$; $\fsvect(m|n)=(\id, \fsl(m|n))_*$;
$\fh(2m|n)=(\id, \fosp^{sk}(m|2n))_*$; $\fle(n)=(\id, \fpe^{sk}(n))_*$;
$\fs\fle(n)=(\id, \fspe^{sk}(n))_*$. 

\begin{rem*}{Remark} Observe that the Cartan prolongs $(\id, \fosp^{sy} (m|2n))_*$
and $(\id, \fpe ^{sy} (n))_*$ are finite dimensional. 
\end{rem*}

The generalization of Cartan's prolongations described in sec. 0.5 has,
after superization, two analogs associated with the contact series $\fk$ and
$\fm$, respectively. 

$\bullet$ First, we define $\fhei(2n|m)$ on the direct sum of a $(2n,
m)$-dimensional superspace $W$ endowed with a nondegenerate skew-symmetric
bilinear form and a $(1, 0)$-dimensional space spanned by $z$.

Clearly, we have $\fk(2n+1|m)=(\fhei(2n|m), \fc(\fosp^{sk}(m|2n)))_*$ and, given
$\fhei(2n|m)$ and a subalgebra $\fg$ of $ \fc(\fosp^{sk}(m|2n))$, we call
$(\fhei(2n|m), \fg)_*$ the {\it $k$-prolong} of $(W, \fg)$, where $W$ is the
identity $ \fc(\fosp^{sk}(m|2n))$-module.

$\bullet$ The odd analog of $\fk$ is associated with the following odd analog of
$\fhei(2n|m)$. Denote by $\fab(n)$ the {\it antibracket} Lie superalgebra: its
space is $W\oplus {\Cee}\cdot z$, where $W$ is an $n|n$-dimensional superspace
endowed with a nondegenerate skew-symmetric odd bilinear form $B$; the
bracket in $\fab(n)$ is given by the following formulas: $z$ is odd and lies in
the center; $[v, w]=B(v, w)\cdot z$ for $v, w\in W$.

Set $\fm(n)=(\fab(n), \fc(\fpe^{sk}(n)))_*$ and, given $\fab(n)$ and
a subalgebra $\fg$ of $\fc(\fpe^{sk}(n))$, we call $(\fab(n),
\fg)_*$ the {\it $m$-prolong} of $(W, \fg)$, where $W$ is the identity
$\fc(\fpe^{sk}(n))$-module.

Generally, given a nondegenerate form $B$ on a superspace $W$ and a
superalgebra $\fg$ that preserves $B$, we refer to the above generalized
prolongations as to {\it $mk$-prolongation} of the pair $(W, \fg)$. 

{\bf Partial Cartan prolong}. The usual Cartan prolongation
starts with nonpositive elements. Define the {\it Cartan prolongation of a positive part}.
Take a
$\fg_0$-submodule
$\fh_1$ in $\fg_1$. Suppose that $[\fg_{-1}, \fh_1]$ is the whole $\fg_0$, not a subalgebra
of $\fg_0$. Define the 2nd prolongation of 
$(\mathop{\oplus}\limits_{i\leq 0}\fg_i, \fh_1)$ to be $\fh_{2}=\{D\in\fg_{2}: [D,
\fg_{-1}]\in
\fh_1\}$. The terms $\fh_{i}$ are similarly defined. Set $\fh_i=\fg_i$ for
$i<0$ and $\fh_*=\sum\fh_i$.

{\it Examples}: $\fvect(1|n; n)$ is a subalgebra of
$\fk(1|2n; n)$. The former is obtained as Cartan's prolong of the same nonpositive part
as $\fk(1|2n; n)$ and a submodule of $\fk(1|2n; n)_1$. The simple exceptional superalgebra
$\fk\fas$ introduced in \S 3 is another example.

\ssec{0.7. Deformations of the Buttin superalgebras and $\fvect(m|n)$-modules} 
Here we reproduce a result of Kotchetkoff [Ko1] with corrections from [Ko2], [L3] and
[LSh]. To consider the deformations,
recall the definition of the
$\fvect(m|n)$-module of tensor fields of type $V$, see [BL]. Let $V$ be the
$\fgl(m|n)= \fvect_0(m|n)$-module with the lowest weight $\lambda$. Make
$V$ into a $\fg_{\geq}$-module, where
$\fg=\fvect(m|n)$ and $\fg_{\geq}=\mathop{\oplus}\limits_{i\geq
0}\fg_{i}$, setting
$\fg_{+}\cdot V=0$ for $\fg_{+}=\mathop{\oplus}\limits_{i>
0}\fg_{i}$. Let us realize $\fg$ by vector fields on the
$m|n$-dimensional linear supermanifold $\cC^{m|n}$ with coordinates $x$. The
superspace $T(V)=\Hom_{U(\fg_{+})}(U(\fg), V)$ is isomorphic, due to the
Poincar\'e--Birkhoff--Witt theorem, to ${\Cee}[[x]]\otimes V$. Its elements
have a natural interpretation as formal tensor fields of type $V$. When $\lambda=(a,
\dots , a)$ we will simply write $T(\vec a)$ instead of $T(\lambda)$.

Examples: $T(\vec 0)$ is the superspace of functions; $\Vol(m|n)=T(1, \dots , 1;
-1, \dots , -1)$ (the semicolon separates the first $m$ coordinates of the
weight with respect to the matrix units $E_{ii}$ of $\fgl(m|n)$) is the
superspace of {\it densities} or {\it volume forms}. We denote the generator
of $\Vol(m|n)$ corresponding to the ordered set of coordinates $x$ by $\vvol(x)$ or
$\vvol_x$. The space of $\lambda$-densities is $\Vol^{\lambda}(m|n)=T(\lambda, \dots ,
\lambda; -\lambda, \dots , -\lambda)$. In particular, $\Vol^{\lambda}(m|0)=T(\vec \lambda)$
but
$\Vol^{\lambda}(0|n)=T(\overrightarrow{-\lambda})$.

As is clear from the definition
of the Buttin bracket, there is a regrading (namely, $\fb (n; n)$ given by
$\deg\xi_i=0, \deg q_i=1$ for all $i$) under which $\fb(n)$, initially
of depth 2, takes the form $\fg=\mathop{\oplus}\limits_{i\geq -1}\fg_i$ with
$\fg_0=\fvect(0|n)$ and
$\fg_{-1}\cong
\Pi(\Cee[\xi])$.

Let us replace the $\fvect(0|n)$-module $\fg_{-1}$ of functions (with inverted
parity) with the module of $\lambda$-densities, i.e., set $\fg_{-1}\cong
\Cee[\xi](\vvol_\xi)^\lambda$, where 
$$
L_D(vol_\xi)^\lambda
=\lambda \Div D\cdot vol_\xi^\lambda\; \text{ and }\;
p(vol_\xi)^\lambda=\od. 
$$
Then the Cartan's prolong $(\fg_{-1}, \fg_0)_*$ is a deform $\fb_{\lambda}(n;
n)$ of $\fb(n; n)$. The collection of these deforms for various $\lambda\in\Cee$
constitutes a deformation of $\fb(n; n)$; we called it the {\it main
deformation}.\index{deformation of the Buttin bracket, main} (Though main, this deformation
is not the quantization of the Buttin bracket, for the latter see [L3].) The deform
$\fb_{\lambda}(n)$ of $\fb(n)$ is the regrading of
$\fb_{\lambda}(n; n)$ inverse to the regrading of $\fb(n)$ into
$\fb(n; n)$.

Another description of the main deformation is as follows. Set 
$$
\fb_{a, b}(n) =\{M_f\in \fm (n):\ a\; \Div M_f=(-1)^{p(f)}2(a-bn)\pderf{f}{\tau}\}.
$$

It is subject to a direct check that $\fb_{a, b}(n)\cong \fb_\lambda(n)$ for
$\lambda =\frac{2a}{n(a-b)}$. This isomorphism shows that $\lambda$ actually
runs over $\Cee\Pee^1$, not $\Cee$. Observe that for $a=nb$, i.e., for
$\lambda=\frac{2}{n-1}$, we have $\fb_{nb, b}(n)\cong \fsm(n)$.

As follows from the description of $\fvect(m|n)$-modules ([BL]) and the
criteria for simplicity of $\Zee$-graded Lie superalgebras ([K]),
the Lie superalgebras $\fb_\lambda(n)$ are simple for
$n>1$ and $\lambda\neq 0, \ -1, \infty$. It is also clear that the $\fb_{\lambda}(n)$
are nonisomorphic for distinct $\lambda$'s. (Notice, that at some values of $\lambda$ the
Lie superalgebras $\fb_{\lambda}(n)$ have additional deformations distinct from the above.
These deformations are partly described in [L3].)

\ssec{0.8. Several first terms that determine the Cartan and
$mk$-prolongations} To facilitate the comparizon of various
vectoral superalgebras, consider the following Table. The central element $z\in\fg_0$ is
supposed to be chosen so that it acts on $\fg_k$ as $k\cdot\id$. The sign $\supplus$
(resp. $\subplus$) denotes the semidirect sum with the subspace or ideal on the left
(right) of it; $\Lambda (r)=\Cee[\xi_1, \dots, \xi_r]$ is the Grassmann superalgebra of the
elements of degree 0.
$$
\renewcommand{\arraystretch}{1.3}
\begin{tabular}{|c|c|c|c|}
\hline
$\fg$&$\fg_{-2}$&$\fg_{-1}$&$\fg_0$\cr
\hline
\hline
$\fvect(n|m; r)$&$-$&$\id\otimes\Lambda (r)$&$\fgl(n|m-r)\otimes\Lambda
(r)\supplus\fvect(0|r)$\cr  
\hline
\hline
$\fsvect(n|m; r)$&$-$&$\id\otimes\Lambda (r)$&$\fsl(n|m-r)\otimes\Lambda
(r)\supplus\fvect(0|r)$\cr  
\hline
$\fvect(1|m; m)$&$-$&$\Lambda (m)$&$\Lambda (m)\supplus\fvect(0|m)$\cr  
\hline
$\fsvect(1|m; m)$&$-$&$\Vol (0|m)$&$\fvect(0|m)$\cr  
\hline
$\fsvect\degree(1|m; m)$&$-$&$v\in\Vol (0|m):\int v=0$&$\fvect(0|m)$\cr  
\hline
$\fsvect\degree(1|2)$&$-$&$T^0(0)\cong\Lambda
(2)/\Cee\cdot 1$&$\fvect(0|2)\cong\fsl(1|2)$\cr    
\hline
$\fsvect(2|1)$&$-$&$\Pi(T^0(0))$&$\fvect(0|2)\cong\fsl(2|1)$\cr  
\hline
\hline
$\fk(2n+1|m; r)$&$\Lambda(r)$&$\id\otimes\Lambda
(r)$&$\fc\fosp(m-2r|2n)\otimes\Lambda (r)\supplus\fvect(0|r)$\cr 
\hline
$\fh(2n|m; r)$&$\Lambda(r)/\Cee\cdot 1$&$\id\otimes\Lambda
(r)$&$\fosp(m-2r|2n)\otimes\Lambda (r)\supplus\fvect(0|r)$\cr 
\hline
$\fk(1|2m; m)$&$-$&$\Lambda(m)$&$ \Lambda (m)\supplus\fvect(0|m)$\cr 
\hline
$\fk(1|2m+1; m)$&$\Lambda(m)$&$\Pi(\Lambda
(m))$&$\Lambda (m)\supplus\fvect(0|m)$\cr 
\hline
\end{tabular} 
$$
Recall that $\fb_{a, b}(n)\cong \fb_\lambda(n)$ for
$\lambda =\frac{2a}{n(a-b)}$.
$$
\renewcommand{\arraystretch}{1.3}
\begin{tabular}{|c|c|c|c|}
\hline
$\fg$&$\fg_{-2}$&$\fg_{-1}$&$\fg_0$\cr
\hline
\hline
$\fb_{\lambda}(n;
r)$&$\Pi(\Lambda(r))$&$\id\otimes\Lambda(r)$&$(\fspe(n-r)\supplus\Cee(az+bd))\otimes\Lambda
(r)\supplus\fvect(0|r)$\cr 
\hline
$\fb_{\lambda}(n; n)$&$-$&$\Pi(\Vol^{\lambda}(0|n))$&$\fvect(0|n)$\cr  
\hline
\hline
$\fm(n; r)$&$\Pi(\Lambda(r))$&$\id\otimes\Lambda(r)$&$\fc\fpe(n-r)\otimes\Lambda
(r)\supplus\fvect(0|r)$\cr 
\hline
$\fm(n; n)$&$-$&$\Pi(\Lambda(n))$&$ \Lambda
(n)\supplus\fvect(0|n)$\cr  
\hline
$\fsb\degree_{\lambda}(n; n)$&$-$
&$\frac{\Pi(\Vol(0|n))}{\Cee(1+\lambda\xi_1\dots\xi_n)\vvol(\xi)}$&
$\fsvect_\lambda (0|n)$\cr 
\hline
\hline
$\fle(n; r)$&$\Pi(\Lambda(r))/\Cee\cdot 1$&$\id\otimes\Lambda
(r)$&$\fpe(n-r)\otimes\Lambda (r)\supplus\fvect(0|r)$\cr 
\hline
\hline
$\fle(n; n)$&$-$&$\Pi(\Lambda(n))/\Cee\cdot 1$&
$\fvect (0|n)$\cr 
\hline
$\fsle\degree(n; r)$&$\Pi(\Lambda(r))/\Cee\cdot 1$&$\id\otimes\Lambda
(r)$&$\fspe(n-r)\otimes\Lambda (r)\supplus\fvect(0|r)$\cr 
\hline
$\fsle\degree(n; n)$&$-$&$\Pi(T^0(0))$&$\fsvect(0|n)$\cr 
\hline
\end{tabular} 
$$

\section*{\S 1. The exceptional Lie superalgebra $(\spin, \fas)_*$ }

\ssec{1.1. A. Sergeev's extension} Let $\omega$ be a nondegenerate
superskew-symmetric bilinear form on an $(n, n)$-dimensional superspace $V$. In
the standard basis of $V$ (all the even vectors come first) the canonical
matrix of the form
$\omega$ is $\begin{pmatrix} 
0&1_n\\ 
1_n&0\end{pmatrix}$ and the elements of 
$\fpe(n)=\faut (\omega)$ can be represented by supermatrices of the form
$\begin{pmatrix}
a & b \cr 
c & -a^t \end{pmatrix}, \text{ where } b=b^t, \; c=-c^t$.
The Lie superalgebra $\fspe(n)$ is
singled out by the requirement that $\tr a=0$.
Setting 
$$ 
\deg \left(\begin{pmatrix}
0 & 0 \cr 
c & 0 \end{pmatrix}\right )=-1,\quad \deg \left(\begin{pmatrix}
a & 0 \cr 
0 & -a^t \end{pmatrix}\right )=0,\quad \deg \left(\begin{pmatrix}
0 & b \cr 
0 & 0 \end{pmatrix}\right )=1,\eqno{(1.1)}
$$
we endow $\fpe(n)$ with a $\Zee$-grading. It is known ([K]) that
$\fspe(n)=\fpe(n)\cap\fsl(n|n)$ is a simple Lie superalgebra for $n\geq 3$. 

A.~Sergeev proved that there is just one nontrivial central extensions of
$\fspe(n)$. It exists for $n=4$ and is denoted by $\fas$. Let
us represent an arbitrary element $A\in\fas$ as a pair $A=x+d\cdot z$, where
$x\in\fspe(4)$,
$d\in{\Cee}$ and $z$ is the central element. In the
matrix form the bracket in $\fas$ is 
$$
\left[\begin{pmatrix} a & b \cr 
c & -a^t \end{pmatrix}+d\cdot z, \begin{pmatrix}
a' & b' \cr 
c' & -a'{}^t \end{pmatrix} +d'\cdot z\right]=\left[\begin{pmatrix} a & b \cr 
c & -a^t \end{pmatrix}, \begin{pmatrix}
a' & b' \cr 
c' & -a'{}^t \end{pmatrix}\right]+\tr~cc'\cdot z.
$$
Clearly, $\deg z= -2$ with respect to the grading  (1.1).

\ssec{1.2} The Lie superalgebra $\fas$ can also be described with the help of
the spinor representation. Consider $\fpo(0|6)$, the Lie superalgebra whose
superspace is the Grassmann superalgebra $\Lambda(\xi, \eta)$ generated by $\xi_1,
\xi_2,
\xi_3, \eta _1, \eta _2, \eta_3$ and the bracket is the Poisson bracket.
Recall that $\fh(0|6)=\Span (H_f: f\in\Lambda (\xi, \eta))$.

Now, observe that $\fspe(4)$ can be embedded into $\fh(0|6)$. Indeed, setting
$\deg \xi_i=\deg \eta _i=1$ for all $i$ we introduce a $\Zee$-grading on 
$\Lambda(\xi, \eta)$ which, in turn, induces a $\Zee$-grading on 
$\fh(0|6)$ of the form $\fh(0|6)=\mathop{\oplus}\limits_{i\geq -1}\fh(0|6)_i$.
Since $\fsl(4)\cong\fo(6)$, we can identify
$\fspe(4)_0$ with $\fh(0|6)_0$. 

It is not difficult to see that the elements
of degree $-1$ in $\fspe(4)$ and $\fh(0|6)$ constitute isomorphic
$\fsl(4)\cong\fo(6)$-modules. It is subject to a direct verification that it is
possible to embed $\fspe(4)_1$ into $\fh(0|6)_1$.

Sergeev's extension $\fas$ is the result of the restriction to
$\fspe(4)\subset\fh(0|6)$ of the cocycle that turns $\fh(0|6)$ into
$\fpo(0|6)$. The quantization deforms $\fpo(0|6)$ into $\fgl(\Lambda(\xi))$;
the through maps
$T_\lambda: \fas\tto\fpo(0|6)\tto\fgl(\Lambda (\xi))$ are representations of
$\fas$ in the
$4|4$-dimensional modules $\spin_\lambda$ isomorphic to each other
for all
$\lambda\neq 0$. (Here
$\lambda\in{\Cee}$ plays the role of Planck's constant.) The explicit
form of
$T_\lambda$ is as follows: 
$$
T_\lambda: \begin{pmatrix} a & b \cr 
c & -a^t \end{pmatrix}+d\cdot z\mapsto \begin{pmatrix}
a & b-\lambda \tilde c \cr 
c & -a^t \end{pmatrix}+\lambda d\cdot 1_{4|4}, \eqno{(1.2)}
$$
where $1_{4|4}$ is the unit matrix and for a skew-symmetric matrix
$c_{ij}=E_{ij}-E_{ji}$ we set $\tilde c_{ij}=c_{kl}$ for the even permutation
$(1234)\mapsto(ijkl)$. Clearly, $T_\lambda$ is an irreducible representation.

\ssbegin{1.3}{Theorem} {\em 1)} The Cartan prolong
$\ff^\lambda=(\spin_\lambda, \fa\fs)_*$ is infinite dimensional and simple for 
$\lambda\neq 0$.

{\em 2)} $\ff^\lambda\cong \ff^\mu$ if $\lambda\cdot \mu\neq 0$.
\end{Theorem} 
 
\begin{Convention} For brevity, we denote the isomorphic
superalgebras $\ff^\lambda=(\spin_\lambda,
\fa\fs)_*$ for any
$\lambda\neq 0$ by $(\spin, \fas)_*$.
\end{Convention}

\begin{proof} Heading 1) consists of two statements: a)
$(\spin, \fas)_k\neq 0$ for all
$k>0$; b) the Lie superalgebra $(\spin_\lambda, \fa\fs)_*$ has no nontrivial $\Zee $-graded
ideals.

a) This follows from the fact that the elements $u_i^{k+1}\partial {\xi_{i}}$
belong to $(\spin, \fas)_k$ for any $k>0$ and any $i$ (to prove the statement it suffices
to consider only one $i$).

b) Assume the contrary: let $\fii=\mathop{\oplus}\limits_{k\geq -1}\fii_k$ be a nonzero
ideal of $\fh=(\spin, \fas)_*$. Let $x\in\fii_k$ be a nonzero homogeneous element.
Since
$\fg=\fvect(4|4)\supset (\spin, \fas)_*$ is transitive, then the superspace ($k+1$
brackets) 
$$
[\fh_{-1}, [\fh_{-1}, \dots , [\fh_{-1}, x]\dots ]=
[\fg_{-1}, [\fg_{-1}, \dots , [\fg_{-1}, x]\dots ]\subset \fii _{-1}
$$ is a nonzero subspace of $\fh_{-1}$. Since $T_\lambda$ is irreducible, 
$\fii _{-1}=\fh_{-1}$. The Jacobi identity implies that $[\fii
_{-1}, \fh_{1}]\subset \fii_0$ is an ideal of $\fh_0$. 

But $\fh_0=\fas$ has only one nontrivial ideal, the center. Since $[\fii
_{-1}, \fh_{1}]=[\fh_{-1}, \fh_{1}]$ contains elements of the form
$u_i\partial {\xi_{i}}$ for any $i$, which do not belong to the center, it
follows that $\fii_0=\fh_0$. In particular, $\fii_0$ contains the element 
$T_\lambda(z)=-\lambda\sum(u_i\partial {u_{i}}+\xi_{i}\partial {\xi_{i}})$. 

But $[T_\lambda(z), h]=-\lambda\cdot k\cdot h$ for any $h\in\fh_k$. Hence,
$\fii=\fh$ and $\fh$ is simple. 

2) is clear.
\end{proof}

\ssec{1.4} For $\lambda=0$ the representation $T_0$ is not faithful and
$T_0(\fas)=\fspe(4)$. The Cartan prolong of the pair $(\id, \fspe(4))$ is
well-known: this is $\fsle(4)$. Recall that we can realize $\fle(n)$ by
the generating functions --- the elements of $\Cee[u, \xi]$ --- with the {\it
Buttin} bracket.
The subalgebra $\fsle(n)$ is generated by functions that satisfy $\Delta
(f)=0$, where $\Delta =\sum_{i=1}^n\frac{\partial^2}{\partial
u_{i}\partial \xi_{i}}$. The exceptional Lie superalgebra $(\spin, \fas)_*$ is a deform of
$\fsle(4)\supplus\Cee\cdot\sum (u_i\partial {u_{i}}+
\xi_i\partial {\xi_{i}})$. An
explicit expression of the corresponding cocycle is desirable: it will enable us to express
the bracket in $(\spin, \fas)_*$ in terms of harmonic functions (plus one more
element).

\section*{\protect\S 2. An explicit form of the vector fields from
\protect $(\spin, \fas)_*\subset\fs\fvect(4|4)$}

Every element $D\in\fvect(4|4)$ is of the form $D=\sum_{i\leq
4}(P_i\partial {\xi_{i}}+Q_i\partial {u_{i}})$, where $P_i, Q_i\in \Cee[u,
\xi]$.

\ssbegin{2.1}{Lemma} The homogeneous (wrt parity) vector field $D\in\fvect(4|4)$ belongs to
$(\spin, \fas)_*$ if and only if it satisfies the following system of equations:
$$
\pderf{Q_i}{u_j}+(-1)^{p(D)}\pderf{P_j}{\xi_i}=0 \text{ for any }i\neq
j;\eqno{(2.1)}
$$
$$
\pderf{Q_i}{u_i}+(-1)^{p(D)}\pderf{P_i}{\xi_i}=\frac{1}{2}\sum_{1\leq j\leq 4}
\pderf{Q_j}{u_j}\text{ for }i=1, 2, 3, 4;\eqno{(2.2)}
 $$
$$
\pderf{Q_i}{\xi_j}+\pderf{Q_j}{\xi_i}=0 \text{ for any }i, j;\eqno{(2.3)}
$$

$$
\pderf{P_i}{u_j}-\pderf{P_j}{u_i}=
(-1)^{p(D)}\cdot\lambda\cdot\big(\pderf{Q_k}{\xi_l}-\pderf{Q_l}{\xi_k}\big)\; \text{for
any even permutation }
 \begin{pmatrix} 1& 2& 3& 4\\ i& j& k& l\end{pmatrix}.
\eqno{(2.4)} 
$$

\end{Lemma}

\ssbegin{2.1.1}{Remark} 1) Observe that the sum of the 4 equations (2.2) yields that
$\Div D=0$, i.e., $(\spin, \fas)_*\subset\fsvect(4|4)$. 

2) For $\lambda =0$ the system (2.1)--(2.4) singles out the
superalgebra 
$$
(\fsle(4)\supplus\Cee\cdot\sum (u_i\partial {u_{i}}+
\xi_i\partial {\xi_{i}}))\supplus\Cee\cdot\Le_{\xi_1\xi_2\xi_3\xi_4}.
$$
\end{Remark}

\ssbegin{2.1.2}{Remark} 1) Actually, ANY Cartan prolongation is obtained as a solution of
some system of differential equations with constant coefficients. In particular,
any $\fg_0\subset\fgl(\fg_{-1})$ is a solution of some homogeneous
linear system $HLS(\fg)$. Interesting examples are given by Lemmas 2.1 and 5.1. Observe
that the number of equations needed to single out
$\fvect(V)$ as a subalgebra in $\fsl(\Lambda(V)/\Cee\cdot 1)$ grows with $\dim V$, since
$\codim
\fvect(V)=(2^{2n}-2^{n+1}) - n2^{n}$. For $\dim V=2$ there are no equations; for $\dim
V=3$ there are 16 equations, etc. Perhaps, these equations can be written in a compact way;
e.g., $\Div D=0$ is a shorthand that eliminates $2^{n}$ parameters from the initial
$n2^{n}$ ones. 

2) Let $\fg_{-1}=V=\Span(\partial_i, i\in I)$. Then any vector
field $D=\sum f_iD_i$ (each $f_i$ is a function on $V^*$) generates a linear operatior
$L_D: V\longrightarrow \fvect(V)$ (Lie derivative): $L_D(D_i)=[D, D_i]$. This operator is a
tensor object  determined by the matrix
$P(D)=(P_{i,j})$, where $P_{i,j} =(-1)^{p(D)p(x_j)}\pderf{f_i}{x_j}$. If $D\in
vect(V)_0$, then the matrix $P(D)$ is a numerical one and if $D\in \fg$, then $D$
satisfies $HLS(\fg)$. Lemma 2.1 proves (but not explaines) that for any $D\in (V, \fg)_*$
the matrix P(D) satisfies $HLS(\fg)$. (The explanation of this fact is very simple:
the LEFT $\ad(D_i)$ commutes with the RIGHT $\ad(D_j)$.) This means that for any $D\in
(V,\fg)_*$ the matrix $P(D)$ is a $\fg$-valued function on $V$! So to describe the Cartan
prolongation $(V, \fg)_*$ we should find out the functions which can be integrated to
vector  fields. 
\end{Remark}

\begin {proof} Denote by $\fg^{\lambda}=\mathop{\oplus}\limits_{i\geq -1}\fg^{\lambda}_i$
for
$\lambda \neq 0$ the space of solutions of the system (2.1)--(2.4). Clearly,
$\fg^{\lambda}_{-1}\cong \fvect(4|4)_{-1}$. Let $D\in \fg^{\lambda}_0$. Then the
matrix of the operator $D$ in its action on $\fg^{\lambda}_{-1}$ is of the form
$$
\begin{pmatrix}a&b\\c&d\end{pmatrix},\; \text{ where }\;
a_{ij}=\pderf{Q_i}{u_j}, b_{ij}=\pderf{P_i}{u_j},
c_{ij}=(-1)^{p(D)}\pderf{Q_i}{\xi_j}, d_{ij}=(-1)^{p(D)}\pderf{P_i}{\xi_j}.
$$
Therefore, equations (2.1) and (2.2) mean that $a+d^t=(\frac 12\tr a)\cdot
1_4$,  equations (2.3) that $c+c^t=0$, equations (2.4) that $b-b^t=\lambda(\tilde
c-\tilde c^t)$. Set
$$
a_0=a-(\frac 14\tr a)\cdot 1_4,\quad d_0=d- (\frac 14\tr a)\cdot 1_4.
$$
Then
$$
a_0+d_0^t=a+d^t-(\frac 12\tr a)\cdot 1_4=0;\quad c+c^t=0;\quad b-b^t=\lambda(\tilde
c-\tilde c^t).\eqno{(2.5)}
$$
Comparing formulas $(2.5)$ with $(1.2)$ we see that $\fg^{\lambda}_0$
coincides with the image of $\fa\fs$ under $T_\lambda$, i.e., with
$(\spin, \fas)_0$.

Set 
$$
D_{u_{j}}(D)=\sum\limits_{i\leq
4}(\pderf{P_i}{u_j}\pder{\xi_i}+\pderf{Q_i}{u_j}\pder{u_i})\text{ and }
\tilde D_{\xi_{j}}(D)=(-1)^{p(D)}\sum\limits_{i\leq
4}(\pderf{P_i}{\xi_j}\pder{\xi_i}+\pderf{Q_i}{\xi_j}\pder{u_i}).
$$
The operators $D_{u_{j}}$ and $\tilde D_{\xi_{j}}$, clearly, {\it commute} with the
$\fg^{\lambda}_{-1}$-action. Observe: the operators {\it commute}, not {\it
super}commute.

Since equations (2.1)--(2.4) is a linear combination of only these operators, the
definition of Cartan prolongation itself ensures an isomorphism of
$(\fg^{\lambda})_{n}$ with $(\ff^\lambda)_{n}$.
\end {proof}

\ssec{2.2. The right inverse of $\Delta$} Let $f$ be an arbitrary homogeneous wrt the
degree in
$u$ and $\xi$ harmonic function, distinct from $\xi_1\dots\xi_n$, i.e., an
arbitrary generating function for $\fsle\degree (n)$. Then $f=\Delta (F)$ for
some function $F$ (as follows from the computation of the homology of $\Delta$ which is an
easy exersise; the answer: the homology space $H(\Delta)$ is spanned by $\xi_1\dots\xi_n$).
Clearly,
$F$ is determined up to an arbitrary harmonic summand. Set $\Phi=\sum u_i\xi_i$. Then
$$
\Delta (\Phi f)=(\Delta\Phi)f-(-1)^{p(f)}\Phi\Delta f-(-1)^{p(f)}\{\Phi, f\}
=\left(n+\deg_u f-\deg_\xi f\right)f.
$$
Define the {\it right inverse} of $\Delta$ by the formula 
$$
\Delta^{-1}f=\frac{1}{\nu(f)}(\Phi f),\; \text{ where }\; \nu(f)=n+\deg_u
f-\deg_\xi f.\eqno{(2.6)}
$$
Since the kernel of $\Delta$ is nonzero, $\Delta$ has no inverse. Still, $\Delta^{-1}$
maps $\fsle\degree (n)$ into $\fle (n)$ without kernel and the following formula
holds:
$$
\Delta(\Delta^{-1}f)=f.
$$

\ssbegin{2.3}{Theorem} Any vector field $D\in\fg^\lambda$ can be represented in the
form
$$
D=D_f+c Z, \; \text{ where }\; c\in\Cee\; \text{ and }\; 
Z=\sum\limits_{i\leq 4}(u_i\partial{u_i}+\xi_i\partial{\xi_i}),
$$
where $f\in\fsle\degree (4)$ and where (recall that $A_n\subset
S_n$ denotes the subgroup of even permutations):
$$
D_f=\Le_f+\lambda\left(-\Le_{\hat{f}}+2\sum\limits_{1\leq
i\leq 4;\;  (i, j, k, l)\in A_{4}}\frac{\partial^{3}(\Delta^{-1}(f))}
{\partial {\xi_{j}}\partial {\xi_{k}}\partial {\xi_{l}}}\partial
{\xi_{i}}\right)\;
\text{for}
\;
\hat{f}= 4\Delta^{-1}\left(\frac{\partial^{4}(\Delta^{-1}(f))}
{\partial {\xi_{1}}\partial {\xi_{2}}\partial {\xi_{3}}\partial {\xi_{4}}}\right ).
$$
\end{Theorem}

For the proof see Appendix 1.

\begin{Corollary} $1)$ The Lie superalgebra $\fg^\lambda$ is a deformation
of the Lie superalgebra $\fsle\degree (4)\supplus \Cee\cdot Z$.

$2)$ If $\deg_\xi f \leq 1$, then $D_f=\Le_f$, hence, $\fh=\{c\cdot
Z+D_f:\deg_\xi f \leq 1, c\in \Cee\}$ remains rigit under this
deformation.

$3)$ Let $\Omega=du_1\wedge du_2\wedge du_3\wedge du_4$ be the volume
element on the underlying manifold of the $\cC^{4|4}$. Observe that the
volume element $\vvol(u, \xi)$ on the whole $\cC^{4|4}$ is invariant wrt the
$\fg^\lambda$-action, but $\Omega$ is not invariant. It is invariant, however, with
respect to the nondeformed subalgebra $\fh$.

$4)$ Let $D\in \fg^\lambda$; let $L_D$ be the Lie derivative. Denote by
$\nabla=\sum \partial u_{i}\partial \xi_{i}$ the bivector dual to $\omega$.
Observe that the lhs of equations $(2.1)-(2.4)$ determine the coefficients of the
2-form $L_D(\omega)$:

equations $(2.1)$ determine the coefficients of $du_jd\xi_i$;

equations $(2.2)$ determine the coefficients of $du_id\xi_i$;

equations $(2.3)$ determine the coefficients of $d\xi_jd\xi_i$;

equations $(2.4)$ determine the coefficients of $du_jdu_i$.

\noindent The rhs of eqs. $(2.2)$ determines the nonzero coefficients of the form
$\frac12(\sum\pderf{Q_i}{u_i})\omega$, while the rhs of eqs. $(2.4)$ determines
the nonzero coefficients of the form
$\lambda(L_D\Omega)*\nabla$, where $*$ is the convolution of tensors.
\end{Corollary}

Therefore, eqs. (2.1)--(2.4) can be rewritten in the form
$$
L_D\omega=\frac 12(\sum\pderf{Q_i}{u_i})\omega+\lambda(L_D\Omega)*\nabla.
$$

Besides, if we replace the rhs of eqs. (2.2) with an arbitrary function $\Psi(u,
\xi)$ but add the constraint
$$
\Div (D)=\sum(\pderf{Q_i}{u_i}-(-1)^{p(D)}\pderf{P_i}{\xi_i})=0,\eqno{(2.7)}
$$
then the sum of the four eqs. (2.2) with eq. $(2.5)$ automatically yields
$$
\Psi(u, \xi)=\frac 12\sum\pderf{Q_i}{u_i}.
$$

Thus, we can distinguish the Lie superalgebra $\fg^\lambda$ by eqs.
$$
\left\{\begin{matrix}
L_D\omega&=&\Psi\cdot\omega&+\lambda(L_D\Omega)*\nabla,\\
\Div (D)&=&0.&
\end{matrix}\right.
$$

\section*{\protect \S 3. The exceptional Lie subsuperalgebra \protect
$\fk\fas$ of \protect $\fk
(1|6)$}

If the operator $d$ that determines a $\Zee$-grading of the Lie superalgebra
$\fg$  does not belong to $\fg$, we denote the Lie
superalgebra $\fg\oplus {\Cee}\cdot d$ by $\fdg$. Recall also that $\fc(\fg)$ or
just $\fc\fg$ denotes the trivial 1-dimensional central extension of $\fg$ with the even
center.

\ssec{3.1} The Lie superalgebra $\fg=\fk(1|2n)$ is generated by the
functions from $\Cee[t, \xi_1, \dots, \xi_n, \eta_1, \dots, \xi_n]$. The
standard $\Zee$-grading of $\fg$ is induced by the $\Zee$-grading of
$\Cee[t, \xi, \eta]$ given by $\deg t=2$, $\deg
\xi_i=\deg\eta_i=1$; namely, $\deg K_f=\deg f-2$. Clearly, in this grading $\fg$
is of depth 2. Let us consider the functions that generate several first
homogeneous components of
$\fg=\mathop{\oplus}\limits_{i\geq -2}\fg_i$: 
$$
\begin{array}{|c|c|c|c|c|}
\hline
\text{component}&\fg_{-2}&\fg_{-1}&\fg_{0}&\fg_{1}\\
\hline
\text{its generators}&1&\Lambda^1(\xi, \eta)&\Lambda^2(\xi,
\eta)\oplus\Cee\cdot t&\Lambda^3(\xi,
\eta)\oplus t\Lambda^1(\xi, \eta)\\
\hline
\end{array}
$$
As one can prove directly, the component $\fg_1$ generates the whole subalgebra
$\fg_+\subset\fg$ of elements of positive degree. The component $\fg_1$ splits into two
$\fg_0$-modules
$\fg_{11}= \Lambda^3$ and 
$\fg_{12}=t\Lambda^1$. It is obvious that $\fg_{12}$ is
always irreducible and the component $\fg_{11}$ is
trivial for $n=1$.

The Cartan prolongations of these components are well-known:
$$
\begin{matrix}
(\fg_-\oplus\fg_0, \fg_{11})_*^{mk}\cong\fpo(0|2n)\oplus\Cee\cdot
K_t\cong\fd(\fpo(0|2n));\\
(\fg_-\oplus\fg_0,
\fg_{12})_*^{mk}=\fg_{-2}\oplus\fg_{-1}\oplus\fg_{0}
\oplus\fg_{12}\oplus\Cee\cdot
K_{t^{2}}\cong\fosp(2n|2).\end{matrix}
$$
Observe a remarkable property of $\fk(1|6)$. For $n>1$ and $n\neq 3$ the component
$\fg_{11}$ is irreducible. For $n=3$ it splits into 2 irreducible conjugate
modules that we will denote by $\fg_{11}^\xi$ and $\fg_{11}^\eta$. Observe
further, that $\fg_0=\fo(6)\cong\fsl(4)$. As $\fsl(4)$-modules,
$\fg_{11}^\xi$ and $\fg_{11}^\eta$ are the symmetric squares $S^2(\id)$
and $S^2(\id^*)$ of the identity 4-dimensional representation and its dual,
respectively.

\ssbegin{3.2}{Theorem} {\em 1)} The Cartan prolong
$(\fg_-\oplus\fg_0, \fg_{11}^\xi\oplus\fg_{12})_*^{mk}$ is infinite
dimensional and simple. It is isomorphic to $(\fg_-\oplus\fg_0,
\fg_{11}^\eta\oplus\fg_{12})_*^{mk}$.

{\em 2)} $(\fg_-\oplus\fg_0, \fg_{11}^\xi)_*^{mk}\cong (\fg_-\oplus\fg_0,
\fg_{11}^\eta)_*^{mk}\cong \fa\fs\oplus\Cee K_t\cong\fd(\fa\fs)$.
\end{Theorem} 
 
\begin{proof} Heading 2) is straightforward; the simplicity in heading
1) follows from Kac's criterion [K]. To see that the 
Cartan prolong $(\fg_-\oplus\fg_0, \fg_{11}^\xi\oplus\fg_{12})_*^{mk}$ is infinite
dimensional we need the following Lemma which clinches the proof. \end{proof}

\begin{Lemma} Denote $\fh=(\fg_-\oplus\fg_0,
\fg_{11}^\xi\oplus\fg_{12})_*^{mk}$. Consider the $\Zee$-grading of $\fh$
induced by the standard grading of $\fk(1|6)$. 

For $k>1$ the operator $T_k=(\ad~K_{t^{2}})|_{\fh_{k}}$ determines an isomorphism of
$\fsl(4)$-modules $\fh_k$ and $\fh_{k+2}$. The operator
$T_1=(\ad~ K_{t^{2}})|_{\fg_{11}^{\xi}}$ determines an isomorphism of
$\fg_{11}^{\xi}$ with its image. 
\end{Lemma}

\begin{proof} We easily check that $K_{t^{2}}\in\fh$ and
$$
\ad~ K_{t^{2}}= 2t(t\partial t+E-2),\quad\text{where }
E=\sum(\xi_i\pder{\xi_i}+\eta_i\pder{\eta_i}).
$$
Therefore, $\Ker(\ad~ K_{t^{2}})$ in $\fk(1|6)$ consists of the fields
generated by the functions $f$ such that $\deg_tf+ \deg_\xi f+\deg_\eta f-2=0$,
i.e., $\Ker(\ad~ K_{t^{2}})\cong \fsl(4)\oplus \fg_{12}\oplus \Cee K_{t^{2}}$.

This makes it clear that, first, $\ad~ K_{t^{2}}$ is $\fsl(4)$-invariant;
second, the operators $T_k$ have no kernel for $k>0$.\end{proof}

We will denote the simple exceptional Lie superalgebra described in heading 1) of Theorem
3.2 by $\fk\fas$.

\section*{\protect \S 4. Lie superalgebras \protect $\fc\fg$. The 
exceptional Cartan prolong \protect $(\Pi (T(\vec 0)/{\Cee}\cdot 1), \fcvect (0|3))_*$}

In order to construct these examples we have to recall (see sec. 0.7) that on
the  supermanifold of purely odd dimension the space of volume forms is
$T(\overrightarrow{-1})$ and the space of half-densities is 
$T(\overrightarrow{-1/2})$ (not $T(\overrightarrow{1})$ and $T(\overrightarrow{1/2})$ as on
manifolds).

\ssec{4.1} Let us now describe a construction of several exceptional simple Lie
superalgebras. Let
$\fu=\fvect(m|n)$, let $\fg=(\fu_{-1}, \fg_0)_*$ be a simple Lie subsuperalgebra
of $\fu$. Let, moreover, there exist an element $D\in\fu_0$ that determines an
exterior derivation of $\fg$ and has no kernel on $\fu_+$. 
Let us study the prolong $\tilde\fg=(\fg_{-1}, \fg_0\oplus \Cee D)_*$.

\begin{Lemma} Either $\tilde\fg$ is simple or $\tilde\fg=\fg\oplus
\Cee D$.
\end{Lemma}

\begin{proof} Let $I$ be a nonzero graded ideal of $\tilde\fg$. The
subsuperspace $(\ad~ \fu_{-1})^{k+1} a$ of $\fu_{-1}$ is nonzero for any nonzero
homogeneous element $a\in\fu_k$ and $k\geq 0$. Since $\fg_{-1}=\fu_{-1}$, the ideal
$I$ contains nonzero elements from $\fg_{-1}$; by simplicity of $\fg$ the ideal $I$
contains the whole $\fg$. If, moreover, $[\fg_{-1}, \tilde \fg_1]=\fg_0$, then by
definition of the Cartan prolongation $\tilde\fg=\fg\oplus \Cee D$.

If, instead, $[\fg_{-1}, \tilde \fg_1]=\fg_0\oplus \Cee D$, then $D\in I$ and since
$[D, \fu_+]=\fu_+$, we derive that $I=\fg$. In other words, $\fg$ is simple.
\end{proof}

\ssbegin{4.2}{Example} Take $\fu=\fvect(2^{n-1}|2^{n-1}-1)$. Consider
$\fu_{-1}$ as $\Pi (T(\vec 0)/{\Cee}\cdot 1)$ and set $\fg_{-1}=\fu_{-1}$,
$\fg_0=\fvect(0|n)$. Clearly,
$\fg_{-1}$ is a
$\fg_0$-module. Then $(\fg _{-1}, \fg_{0})_*$ is a simple Lie
superalgebra isomorphic to $\fle(n; n)$. The isomorphism is established with the help of
a regrading. For the operator $D$ of the exterior derivation of $\fle(n; n)$ we take the
grading operator
$d\in
\fm(n; n)_0\subset\fu_0$, i.e.,
$\fg_0\oplus \Cee D\cong\fcvect(0|n)$.

In particular, for $n=2$ we have $\fg_{-1}=\Pi (\xi_1, \xi_2, \xi_1\xi_2)$;
$\fg_0=\fvect(0|2)\cong\fsl(2|1)$. Then $\fc(\fg_{0})=\fgl(2|1)$ and
$(\Pi (T(\vec 0)/{\Cee}\cdot 1), \fcvect (0|2))_*\cong\fvect(2|1)$.
\end{Example}

\begin{Theorem} {\em 1)} $(\Pi (T(\vec 0)/{\Cee}\cdot 1), \fcvect (0|3))_*$ is a simple Lie
superalgebra. 

{\em 2)} $(\Pi (T(\vec 0)/{\Cee}\cdot 1), \fcvect (0|n))_*\cong \fd(\fle(n; n))$ for
$n>3$.
\end{Theorem}

\begin{proof} Thanks to Lemma 4.1.1 heading 1) follows from the fact that $(\Pi (T(\vec
0)/{\Cee}\cdot 1), \fcvect (0|3))_*$, which is  the Cartan
prolong, is bigger than $\fs\fle\degree(3; 3)\oplus \Cee
D$; we will prove this fact in \S 5. Heading 2) is proved in Appendix 2.
\end{proof}

\ssec{4.3} To clarify the structure of the
exceptional Lie superalgebra $(\Pi (T(\vec 0)/{\Cee}\cdot 1), \fcvect (0|3))_*$, consider 
one more construction. Let us  describe one wonderful property of $\fs\fle\degree(3)$ that
singles it out among the
$\fs\fle\degree(n)$.

In the standard grading of $\fg=\fs\fle\degree(3)$ we have: $\dim
\fg_{-1}=(3,3)$, $\fg_0\cong\fspe(3)$. For the regraded superalgebra
$R\fg=\fs\fle\degree(3;3)$ we have: $\dim
R\fg_{-1}=(3,3)$, $R\fg_0=\fs\fvect(0|3)\cong\fspe(3)$. Therefore, for 
$\fs\fle\degree(3)$ and only for it among the $\fs\fle\degree(n)$, the regrading $R$
determines a nontrivial automorphism. In terms of generating functions the regrading $R$
is given by the formulas:

1) $\deg_\xi(f)=0$: $R(f)=\Delta (f\xi_1\xi_2\xi_3)$;

2) $\deg_\xi(f)=1$: $R(f)=f$;

3) $\deg_\xi(f)=2$:
$R(f)=\frac{\partial^3(\Delta^{-1}f)}{\partial\xi_1\partial\xi_2\partial\xi_3}$
(see (2.6)). 

We see that $R^2=SIGN$, where the operator $SIGN$ is defined by the formulas
$SIGN(D)=(-1)^{p(D)}D$ on the vector fields and $SIGN(f)= (-1)^{p(f)+1}f$
on the generating functions. 

Let now $\fg=\fle(3, 3)$ and $i_1:\fg\tto\fu=\fvect (4|3)$
be the embedding that preserves the standard
$\Zee$-grading of $\fg$. Let $\fh=\fle(3)$ and
$\tilde\fh=\fsle\degree(3)\subset \fh$. Then the map $i_2=SIGN\circ i_1R$ determines an
embedding
$\tilde\fh\tto \fu$ that preserves the grading of $\tilde\fh$.

Observe that $\tilde\fh_0=\fspe(3)$ and
$\fh_0=\fpe(3)\cong\tilde\fh_0\oplus\Cee(\sum q_i\xi_i)$. The action of
$z=-2i_1(\sum q_i\xi_i)+3d$ on the space $i_2(\fh_{-1})$ coincides with the action
of $\sum q_i\xi_i$ on $\fh_{-1}$. Therefore, setting $i_2(\sum q_i\xi_i)=z$ we get
an embedding $i_2(\fh_{-1}\oplus\fh_0)\tto \fu$ that can be extended to an embedding
of $\fh$ to $\fu$. Under this embedding
$$
i_1(\fg_{-1}\oplus \fg_0)+i_2(\fh_{-1}\oplus\fh_0)=\fu_{-1}\oplus(\fg_0\oplus
\Cee d),
$$
i.e., the {\it nondirect} sum of the images of $i_1$ and $i_2$ covers the whole
nonpositive part of
$(\Pi (T(\vec 0)/{\Cee}\cdot 1), \fcvect (0|3))_*$. 

Thus, we obtained two distinct embeddings of $\fle(3)\cong \fle(3;3)$ into $(\Pi (T(\vec
0)/{\Cee}\cdot 1), \fcvect (0|3))_*$:
$$
i_1: \fle(3; 3)\tto(\Pi (T(\vec
0)/{\Cee}\cdot 1), \fcvect (0|3))_*\quad \text{and}\quad i_2: \fle(3)\tto(\Pi (T(\vec 0)/{\Cee}\cdot 1), \fcvect
(0|3))_*\eqno{(4.1)}
$$
such that $i_1(\fle(3; 3))+i_2(\fle(3))=(\Pi (T(\vec
0)/{\Cee}\cdot 1), \fcvect(0|3))_*$ (the sum in the lhs is {\it not}
a direct one!). As a linear space, $(\Pi (T(\vec
0)/{\Cee}\cdot 1), \fcvect (0|3))_*$ is the quotient of $\fle(3;
3))\oplus \fle(3)$ modulo the subspace
$V=\{(SIGN\circ Rg\oplus -g): g\in \fsle\degree(3)\}$. The map $\varphi$ defined by the
formula
$$
\varphi|_{i_1(\fle(3; 3))}=SIGN\circ i_2i_1^{-1};\quad \quad 
\varphi|_{i_2(\fle(3))}=i_1i_2^{-1}
$$
determines a nontrivial automorphism of $(\Pi (T(\vec 0)/{\Cee}\cdot 1), \fcvect (0|3))_*$.

\section*{\protect\S 5. An explicit form of the vector fields from
\protect $(\Pi (T(\vec 0)/{\Cee}\cdot 1), \fcvect (0|3))_*\subset\fvect(4|3)$}

Let us redenote the basis in $\Pi(\Lambda (\eta_1, \eta_2, \eta_3))$ by setting:
$$
\Pi(\eta_1\eta_2\eta_3)\mapsto \partial y; \quad \Pi(\eta_i)\mapsto \partial u_i;\quad 
\Pi(\frac{\partial \eta_1\eta_2\eta_3}{\partial\eta_i})\mapsto \partial \xi_i.
$$
 
Every element $D\in\fvect(4|3)$ is of the form
$D=\sum_{i\leq 3}(P_i\partial {\xi_{i}}+Q_i\partial {u_{i}})+R\partial y$, where
$P_i, Q_i, R\in\Cee[y, u, \xi]$. 

\ssbegin{5.1}{Lemma} Set $\fg_{-1}=\Span(\partial y;
\partial u_i, \partial \xi_i\; \text{for}\; i\leq 3)$, $\fg_{0}=\fcvect (0|3)$. The homogeneous
(wrt parity) vector field $D\in \fvect(4|3)$ belongs to $(\fg_{-1}, \fg_{0})_*$ if and only if
it satisfies the following system of equations:
$$
\pderf{Q_i}{u_j}+(-1)^{p(D)}\pderf{P_j}{\xi_i}=0 \text{ for any }i\neq
j;\eqno{(5.1)}
$$
$$
\pderf{Q_i}{u_i}+(-1)^{p(D)}\pderf{P_i}{\xi_i}=\frac{1}{2}\left(\sum_{1\leq
j\leq 3} \pderf{Q_j}{u_j}+\pderf{R}{y}\right )\text{ for }i=1, 2, 3;\eqno{(5.2)}
 $$
$$
\pderf{Q_i}{\xi_j}+\pderf{Q_j}{\xi_i}=0 \text{ for any }i, j;\text{ in particular
}\pderf{Q_i}{\xi_i}=0;\eqno{(5.3)}
$$
$$
\pderf{P_i}{u_j}-\pderf{P_j}{u_i}=
-(-1)^{p(D)}\pderf{R}{\xi_k}\;
\text{ for any $k$ and any even permutation }
 \begin{pmatrix} 1& 2& 3\\ i& j& k\end{pmatrix};
\eqno{(5.4)} 
$$
$$
\pderf{Q_i}{y}=0\text{ for }i=1, 2, 3;\eqno{(5.5)}
$$
$$
\pderf{P_k}{y}=
(-1)^{p(D)}\frac 12\big(\pderf{Q_i}{\xi_j}-\pderf{Q_j}{\xi_i}\big)\;
\text{ for any $k$ and for any even permutation }
 \begin{pmatrix} 1& 2& 3\\ i& j& k\end{pmatrix}.
\eqno{(5.6)} 
$$
\end{Lemma}

Proof is similar to that of Lemma 2.1.

\begin{rem*}{Remark} The left hand sides of eqs.
(5.1)--(5.6) determine the coefficients of the 2-form $L_D\omega$, where $L_D$ is the
Lie derivative and
$\omega=\sum\limits_{1\leq i\leq 3}du_id\xi_i$. It is interesting to interpret the
rhs of these equations.
\end{rem*}

\ssbegin{5.2}{Theorem} Every solution of the system
$(5.1)-(5.6)$ is of the form:
$$
\begin{matrix}
D=\Le_f+yA_f-(-1)^{p(f)}\left(y\Delta (f)+y^2\frac{\partial^3
f}{\partial \xi_{1}\partial \xi_{2}\partial \xi_{3}}\right )\partial y+\\
A_g-(-1)^{p(g)}\left(y\Delta (g)+2y\frac{\partial^3
g}{\partial \xi_{1}\partial \xi_{2}\partial \xi_{3}}\right)\partial y,
\end{matrix}\eqno{(5.7)}
$$
where $f, g\in\Cee [u, \xi]$ are arbitrary and the operator $A_f$ is given by
the formula:
$$
A_f=\frac{\partial^2f}{\partial\xi_{2}\partial
\xi_{3}}\pder{\xi_{1}}+\frac{\partial^2f}{\partial
\xi_{3}\partial
\xi_{1}}\pder{\xi_{2}}+\frac{\partial^2f}{\partial \xi_{1}\partial
\xi_{2}}\pder{\xi_{3}}.\eqno{(5.8)} 
$$
\end{Theorem} 
 
Proof is similar to that of Theorem 2.3, see Appendix 1 and [PSh].

Formula $(5.7)$ makes it possible to explicitely express the two embeddings
(4.1)
$$
i_1, i_2:\fle(3)\tto(\Pi (T(\vec 0)/{\Cee}\cdot 1), \fcvect (0|3))_*. 
$$

The first embedding $i_1$ preserves the grading of $\fle(3; 3)$, cf. 0.4. I do not
know any compact general formula for $i_1$ and can only determine it
component-wise (mind that $A_3$ in the first line of the following displayed formula is the
group of even permutations, {\it not} the operator $A_f$ for $f=3$):
$$
\begin{array}{rcl}
i_1(\Le_{f(u)})&=&\Le_{\sum\pderf{f}{u_i}\xi_j\xi_k-yf},\;\text{where $y$ is
treated as a parameter}\cr
&&\text{and }(i, j, k)\in A_3;\cr 
i_1(\Le_{\sum f_i(u)\xi_i})&=&\Le_f-\varphi(u)\sum
\xi_i\partial \xi_i+\left(-\varphi(u)y+\Delta
(\varphi(u)\xi_1\xi_2\xi_3)\right)\partial y,\cr
&&\text{where
}\varphi(u)=\Delta(f);\cr 
i_1\left(\Le_{\sum\limits_{1\leq i\leq 3;\; (i, j, k)\in A_3}
\psi_i(u)\xi_k\xi_l}\right)&=&A_f-\Delta(f)\partial y,\;\text{where $A_f$ is
given by (5.8)};\cr
i_1(\Le_{\psi(u)\xi_1\xi_2\xi_3})&=&-\psi(u)\partial y.
\end{array}
$$

The second embedding $i_2$ preserves the standard grading of $\fle(3)$. It is given
by the formulas:
$$
i_2(\Le_f)=\Le_f+yA_f+(-1)^{p(f)}\left(y\Delta (f)+y^2\frac{\partial^3
f}{\partial \xi_{1}\partial \xi_{2}\partial \xi_{3}}\right)\partial y.
$$

\section*{\S 6. Exceptional simple Lie superalgebras of depth 2: \protect
$(\fab (4), \fcvect (0|3))_*^{m}$ and \protect $(\fhei(8|6), \fsvect _{3, 4}(4))_*^{k}$}

Two more examples of exceptional simple Lie superalgebras are obtained with the
help of a construction that generalizes the constructions from \S 4 to Lie superalgebras of
depth 2. Let $\fu=\mathop{\oplus}\limits_{i\geq -2}\fu_i$ be either $\fm(n)$ or
$\fk(2m+1|n)$; let $\fg=(\fu_-, \fg_0)_*$ be a subalgebra of $\fu$ such that the subspace
$\fu_{-2}$ belongs to the center of $\fg$ and the quotient $\fg/\fu_{-2}$ is simple. Let,
moreover,
$D\in\fu_0$ determine an exterior derivation of $\fg$ without kernel on
$\fu_{-2}\oplus\fu_+$, where $\fu_+=\mathop{\oplus}\limits_{i>0}\fu_i$. 

Let us study the $mk$-prolong $\tilde\fg=(\fg_{-}, \fg_0\oplus \Cee D)^{mk}_*$.
The main result of this section: a description of two simple exceptional Lie
superalgebras $(\fab(4), \fcvect (0|3))_*^{m}$ (Th. 6.2) and $(\fhei(8|6), \fsvect _{3,
4}(4))_*^{k}$ (Th. 6.5).

\ssbegin{6.1}{Lemma} Either $\tilde\fg$ is simple or $\tilde\fg\cong\fg\oplus
\Cee D$.
\end{Lemma}

\begin{proof} First, let us prove that if $\tilde\fg\not \cong\fg\oplus \Cee
D$ , then $\fu_{-2}$ is not an ideal in $\tilde\fg$. Indeed, in this case there
exist
$g_{-1}\in\fg_{-1}$, $g_{0}\in\fg_{0}$ and $g_{1}\in\tilde\fg_{1}$ such that 
$[g_{1}, g_{-1}]=D+g_{0}$. 

Let $\fu_{-2}=\Cee z$. Then $[g_{-1}, z]=[g_{0}, z]=0$ and we have
$$
\begin{matrix}
[g_{-1}, [z, g_{1}]]=[[g_{-1}, z], g_{1}]+(-1)^{p(g_{-1})p(z)}[z, [g_{-1},
g_{1}]]\\
(-1)^{p(g_{-1})p(z)}[z, D+
g_{0}]=(-1)^{p(g_{-1})p(z)}[z, D]\neq 0.
\end{matrix}
$$
(We have taken into account that $D$ has no kernel on $\fu_{-2}$.) Hence, $[z,
g_{1}]$ is a nonzero element of $\fg_{-1}$. The rest of the proof mimics that of
Lemma 4.1.1.
\end{proof}

\ssec{6.2} Consider $\hat\fg=\fb_{-1/2}(n; n)$. We have:
$$
\hat\fg_{-1} = \Pi (T(\overrightarrow{-1/2})); \quad 
\hat\fg_{0} = \fcvect (0|n)
$$
and the $\hat\fg_{0}$-action on $\hat\fg_{-1}$ preserves the nondegenerate
superskew-symmetric form
$$
B(\varphi\sqrt{\vvol}, \psi\sqrt{\vvol})=\int \varphi\psi\cdot\vvol; \quad\quad
p(B)\equiv n\pmod 2.\eqno{(6.1)} 
$$

Now, let $\fg=\tilde\fc(\fb_{1/2}(3; 3))$ be the {\it nontrivial} central
extension (we indicate this fact by tilde over $\fc$) of $\hat\fg$ corresponding to
$(6.1)$. The depth of $\tilde\fc(\fb_{1/2}(3; 3))$ is equal to 2. This central extention is
naturally embedded into
$$
\fu=\left\{\begin{matrix}
\fm(2^{n-1}) &\text{for $n$ odd}\\
\fk(1+2^{n-1}|2^{n-1})&\text{for $n$ even}.
\end{matrix}\right .
$$
As the operator $D$ described in sec. 6.1 we take the
grading operator $d\in \fu_0$, i.e., $\fg_0\oplus \Cee D\cong \fc\fg_0$.

\begin{rem*}{Example} Let $n=2$. Then $\fg _{-1} = \Pi
(T(\overrightarrow{-1/2}))$ and $\fg _{-} = \fhei (2|2)$. We also have $\fc\fg _{0} =
\fc\fvect (0|2)\cong\fc\fosp(2|2)$ and $(\fhei (2|2), \fcvect
(0|2))_*^{k} \cong \fk (3|2)$, see sec. 0.6.
\end{rem*}

\begin{Theorem} {\em 1)} $(\fab(4), \fcvect (0|3))_*^{m}$ is a simple
Lie superalgebra. 

{\em 2)} $(\fab(2^{n-1}), \fcvect (0|n))_*^{mk}\cong
\fd((\fab(2^{n-1}), \fvect (0|n))_*^{mk})\cong\fd(\fb(n; n))$ for
$n>3$.
\end{Theorem}

\begin{proof} Thanks to Lemma 6.1 heading 1) follows from the fact that the
$m$-prolongation of $(\fab(4), \fcvect (0|3))$ is bigger than $(\fab(4), \fvect
(0|3))_*^{m}$. We give explicit formulas in Appendix 3.
Heading 2) is proved in Appendix 2.
\end{proof}

\ssec{6.3} Let us clarify the structure of the exceptional Lie superalgebra
$(\fab(4), \fcvect (0|3))_*^{mk}$ with the help of a construction
similar to that from
\S 4. To this end, we describe another remarkable property of
$\fs\fle\degree(3)$ that singles it out among the $\fs\fle\degree(n)$.

The Lie superalgebra $\fg=\fs\fle\degree(3)$ has a $2\varepsilon$-dimensional
nontrivial central extension $\fe\fs\fle\degree(3)$: the element $M_1$ of degree $-2$
with respect to the standard grading of $\fs\fle\degree(3)$ extends
$\fs\fle\degree(3)$ to $\fs\fb\degree(3)$ while $z$ of degree $-2$ with respect to the
grading of $\fs\fle\degree(3; 3)$ is associated with the form $B$ on the space $\fg
_{-1}$ of halfdensities with shifted parity (see (6.1)) in the realization
$\fg=\fs\fle\degree(3; 3)$.

The regrading $R$ interchanges these central elements and establishes a
nontrivial automorphism of $\fe\fs\fle\degree(3)$. 

Let now $\fg=\tilde\fc(\fb_{-1/2}(3; 3))$ be the described in sec. 6.2 nontrivial central
extension of depth 2 of $\fb_{-1/2}(3; 3)$; clearly, $\fg _{-2} =\Cee z$. The
inverse regrading
$R^{-1}$ sends $\fg$ into the 
{\it nontrivial} central extension $\fh=\tilde\fc(\fb_{-3, 1}(3))$ of $\fb_{-3, 1}(3)$,
see sec. 0.7, and $\deg R^{-1}z=-1$.

From the very beginning we have an embedding $i_1:\fg\tto\fu=\fm (4)$. Let
$\tilde\fh=\fe\fsle\degree(3)\subset \fh$. 

Then the map $i_2=SIGN\circ i_1R$ determines an embedding of $\tilde\fh$ to $\fu$
preserving the grading of $\tilde\fh$.

Observe that $\tilde\fh_0=\fspe(3)$ and
$\fh_0=\fpe(3)\cong\tilde\fh_0\oplus\Cee\cdot M_{\sum q_i\xi_i-3\tau}$. 
The action of
$z=M_{\sum q_i\xi_i-3\tau}$ on the space $\fh_{-1}$ is of the form:
$$
M_q\mapsto 4M_q, \quad M_\xi\mapsto 2M_\xi.
$$
This coincides with the action of $-3d+\frac 12i_1(z)$.

Therefore, the embedding $i_2$ of $\fh_{-}$ can be extended to an embedding of the
whole $\fh$. We have
$$
i_1(\fg_{-}\oplus\fg_0)+ i_2(\fh_{-}\oplus\fh_0)=\fu_{-}\oplus(\fg_0\oplus
\Cee d),
$$
i.e., the images of $i_1$ and $i_2$ cover the whole nonpositive part of $(\fab(4), \fcvect
(0|3))_*^{mk}$. 

Thus, we have got two distinct embeddings of $\tilde\fc(\fb_{-3, 1}(3))$,
isomorphic to $\tilde\fc(\fb_{-1/2}(3; 3))$ as abstract, but not as graded, Lie
superalgebras, into $(\fab(4), \fcvect (0|3))_*^{mk}$:
$$
i_1: \tilde\fc(\fb_{-1/2}(3; 3))\tto(\fab(4), \fcvect
(0|3))_*^{mk}\quad \text{with the grading of $\tilde\fc(\fb_{-1/2}(3; 3))$
preserved }
$$
and
$$
i_2: \tilde\fc(\fb_{-3, 1}(3))\tto(\fab(4), \fcvect (0|3))_*^{mk} \quad \text{with the
grading of $\tilde\fc(\fb_{-3, 1}(3))$ preserved }
$$
such that $i_1(\tilde\fc(\fb_{-1/2}(3; 3)))+i_2(\tilde\fc(\fb_{-3,
1}(3)))=(\fab(4), \fcvect(0|3))_*^{mk}$ (the sum here is {\it not} a direct one!). As a
linear space, $(\fcvect (0|3))_*^{mk}$ is the quotient of
$\tilde\fc(\fb_{-1/2}(3; 3))\oplus \tilde\fc(\fb_{-3, 1}(3))$ modulo the subspace
$V=\{(SIGN\circ Rg\oplus -g): g\in \fe\fsle\degree(3)\}$. The map $\varphi$ defined by the
formula
$$
\varphi|_{i_1(\tilde\fc(\fb_{-1/2}(3; 3)))}=SIGN\circ i_2i_1^{-1};\quad \quad 
\varphi|_{i_2(\tilde\fc(\fb_{-3, 1}(3)))}=i_1i_2^{-1}
$$
determines a nontrivial automorphism of $(\fab(4), \fcvect(0|3))_*^{mk}$.

\ssec{6.4. Description of $(\fhei(8|6), \fsvect _{a, b}(n))_*^{mk}$} Consider the nontrivial
central extension
$\fg=\tilde\fc\fsle\degree(n; n)$ of $\fsle\degree(n; n)$ defined as follows.
We have:
$\fg _{0} = \fsvect (0|n)$; $\fg _{-1}=\Pi (T^{0}(\vec 0) /{\Cee}\cdot 1)$,
where $T^{0}(\vec 0) =\{f \in T(\vec 0) :
\int f\vvol(\xi) =0 \}$. Define the central extension with the
help of the form $\omega$ on $\fg _{-1}$ given by the formula:
$$ 
\omega (f, g) = \int fg \cdot\vvol(\xi).
$$ 
The same arguments as in 6.2, show that $(\fg_{-1}, \fg_0)_*^{mk}$ can
be embedded into
$\fk(1+2^{n-1}|2^{n-1}-2)$ for
$n$ even and into $\fm(2^{n-1}-1)$ for $n$ odd. 

Let $x$ be the operator determining the standard $\Zee$-grading of
$\fsvect (0|n)$ and let $z$ commute with $\fsvect (0|n)$; let $a, b \in
\Cee$. For any $a, b$ the element $ax+bz$ determines an outer derivation of
$\fg_0$. Set $\fsvect _{a, b}(n)=\fsvect(0|n)\supplus
\Cee(ax+bz)$; set also 
$$
(\fg_-, \fsvect _{a,
b}(n))_*^{mk}=(\fg_-,
\fg_0\supplus
\Cee(ax+bz))_*^{mk},\; \text{where}\; \fg_-=\left\{\begin{matrix}
\fab(2^{n-1}-1) &\text{for $n$ odd}\\
\fhei(2^{n-1}|2^{n-1}-2)&\text{for $n$ even}.
\end{matrix}\right .
$$ 

\begin{rem*}{Example} Let $n=3$. Then $\fg _{-1} = \Pi
(\xi_1, \xi_2, \xi_3, \xi_1\xi_2, \xi_1\xi_3, \xi_2\xi_3))$ and $\fg _{-}\cong
(\fb_\lambda(3))_-=\fab(3)$ for any $\lambda$; $\fg _{0} = \fsvect (0|3)\cong\fspe(3)$. The
operator $x$ becomes $\sum\xi_i\partial\xi_i$ and 
$$
\fg_0\supplus \Cee(ax+bz)\cong
\fspe(3)\supplus \Cee(a\sum\xi_i\partial\xi_i +bz)\cong (\fb_\lambda(3))_0\; \text{for}\;
\lambda=-\frac{b}{3a}.
$$
Therefore, $(\fab(3), \fsvect _{a, b}(3))_*^{mk}\cong \fb_\lambda(3)$ for
$\lambda=-\frac{b}{3a}$. In particular, $(\fab(3), \fsvect _{1, 3}(3))_*^{mk}\cong
\fs\fm(3)$ and $(\fab(3), \fsvect _{1, 0}(3))_*^{mk}\cong \fb(3)$.
\end{rem*}

\ssbegin{6.5}{Theorem} {\em 1)} $(\fhei(8|6), \fsvect _{3, 4}(4))_*^{mk}$ is a simple Lie
superalgebra. 

{\em 2)} Let $\fg_-=\left\{\begin{matrix}
\fab(2^{n-1}-1) &\text{for $n$ odd}\\
\fhei(2^{n-1}|2^{n-1}-2)&\text{for $n$ even}.
\end{matrix}\right .$ Then 
$$
(\fg _-, \fsvect _{a, b}(n))_*^{mk}\cong (\fg _-, \fsvect
(0|n))_*^{mk}\supplus\Cee(ax+bz)\; \text{if}\; n>4\; \text{or if}\; (a, b)\not\in\Cee(3,
4)\; \text{and}\; n=4.
$$
\end{Theorem}

\begin{proof} As in Theorem 6.1 heading 1) follows from a direct calculations
based on Lemma 6.1; for the explicit formulas see Appendix 3. Heading 2) is proved
in Appendix 2.
\end{proof}

Let us clarify the structure of $(\fhei(8|6), \fsvect _{3, 4}(4))_*^{k}$. This Lie
superalgebra is contained in
$\fu=\fk(9|6)$. In sec. 6.3 we have already described the Lie superalgebra $\fg=
\tilde\fc\fsle\degree(4; 4)$ and its embedding $i_1: \fg\to \fu$. 

Observe that $\fg\supset \fas$ and this embedding preserves the $\Zee$-grading
described in sec. 2.1:
$$
\begin{matrix}
\fas_{-2}=\fg_{-2};\quad\fas_{-1}=\fg_{-1}=\Pi(\Lambda^2(\xi_1,
\xi_2, \xi_3, \xi_4))\\
\fas_{0}=\fsl(4)\subset\fg_0=\fsvect(0|4);\quad\fas_{1}=\Pi(S^2(q_1,
q_2, q_3, q_4))\subset\fg_{1}.
\end{matrix}
$$

For the role of $\fh$ (see 4.3 and 6.3) take
$\fk\fas$. It follows from Theorem 3.2 that $\fas\subset\fk\fas$; set
$\tilde\fh=\fas$. Let $R: \tilde \fh\tto \fg$ be the embedding that executes the
isomorphism of two copies of $\fas$. (Notice that $R$ preserves the $\Zee$-grading (1.1)
of $\fas$.) The map $i_2=i_1R$ determines an embedding of $\tilde\fh$ into $\fk(9|6)$.

But $\fh_0=\tilde\fh_0\oplus \Cee t$. It turns out
that $i_2$ can be extended to an embedding $i_2: \fk\fas\tto \fu$. 

As in the above examples, we have:
$i_1(\fg_{-}\oplus\fg_0)+i_2(\fh_{-}\oplus\fh_0)\cong\fu_{-}\supplus \Cee(3x+4z)$ (the
sum in the lhs here is {\it not} a direct one!). As a linear space, $((\fhei(8|6),
\fsvect_{3, 4} (0|4))_*^{mk}$ is a nondirect sum of $\fg=\tilde\fc(\fsle\degree(4; 4))$ with
$\fh=\fk\fas$ and is the quotient of
$\fg\oplus
\fh$ modulo the subspace
$V=\{(Rg\oplus -g): g\in \fas\}$.

\section*{Appendix 1. Solution of the system $(2.1)$--$(2.4)$} 

Let $D=\sum_{1\leq i\leq 4}(P_i
\pder{\xi_i}+Q_i\pder{u_i})\in\fg^{\lambda}$ be an homogeneous (wrt parity) vector field.
Then by Lemma 2.1 it satisfies the system (2.1)-(2.4).
        
Equations $(2.3)$ imply that there
exists a function $f=f(u, \xi)$ such that $Q_i=-(-1)^{p(D)}\pderf{
f}{\xi_{i}}$. Equations (2.1) imply further that 
$$
P_i=\pderf{f}{{u_{i}}}+\varphi_i(u; \xi_i)=\pderf{
f}{u_{i}}+\varphi_i\degree(u)+\varphi_i^1(u)\cdot \xi_i
$$
or, in other words, $D= Le_{f}+\sum(\varphi_i\degree(u)+\varphi_i^1(u)\xi_i)\pder{\xi_i}$.
Equations $(2.2)$ now imply that
$$
\varphi_i^1(u)=\pderf{\varphi_i}{\xi_{i}}=
\frac12(-1)^{p(D)}\sum\pderf{Q_{j}}{u_{j}}=
-\frac12\Delta (f),\;  \text{ where}\; \Delta=\sum\frac{\partial^2}{\partial {u_{i}}\partial
{\xi_{i}}};\eqno{(2.2')}
$$
whereas equations $(2.4)$ imply that
$$
\pderf{\varphi_i}{u_{j}} -\pderf{\varphi_j}{u_{i}}
=-2\lambda\frac{\partial^2 f}{\partial {\xi_{l}}\partial {\xi_{k}}}.\eqno{(2.4')} 
$$

\begin{rem*}{Remark} Let $\psi_i=\psi_i(u)$, where $i=1,...,4$, be a set of functions
such that $\pderf{\psi_i}{u_j}-\pderf{\psi_j}{u_i}=0$. Then there exists a function $\psi(u)$
for which $\psi_i=\pderf{\psi}{u_i}$ and $\sum\psi_i(u)\pder{\xi_i}=Le_{\psi}=D_{\psi}$
(see Corollary 2 from Theorem 2.3). Thus, for any function $f$ it suffices to find
any collection of functions $\varphi_i\degree$.
\end{rem*}
        
With the help of the differential forms 
$$
\alpha=\sum_{i\leq 4}\varphi_idu_i\text{ and } \omega_0(f)=
\sum_{(i, j, k, l)\in A_{4}}\frac{\partial^2
f}{\partial {\xi_{l}}\partial {\xi_{k}}}du_j\wedge du_i
$$
equations $(2.2')$ and $(2.4')$ can be expressed in the form
$$
d\alpha= \frac{1}{2}\Delta (f)\cdot
\omega-2\lambda\cdot\omega_0(f).\eqno{(A.1)}
$$

Equation (A.1) is solbvable if and only if the form in the rhs is exact or,
since our considerations are local, equivalently, if and only if it is closed.

Direct calculations show that the condition that the form in the rhs is closed
is equivalent to the system
$$
\left\{\begin{matrix}
\pderf{\Delta f}{\xi_i}=0\text{ for all }i= 1, 2, 3, 4\\
\frac 12\pderf{\Delta f}{u_i}-2\lambda\frac{\partial^3
f}{\partial {\xi_{j}}\partial {\xi_{k}}\partial {\xi_{l}}}=0\text{ for all }
i= 1, 2, 3, 4\text{ and }(i, j, k, l)\in A_4
\end{matrix}\right.\eqno{\begin{matrix}(A.1.1)\\(A.1.2)\end{matrix}}
$$

Eqs. $(A.1.1)$ imply that $\Delta (f)$ only depends on $u$; eqs. $(A.1.2)$
imply that $\deg_\xi f\leq 3$. 

First, suppose that $\deg_\xi f\leq 2$. Then eqs. $(A.1.2)$
imply that 
$$
\Delta(f)=const. \eqno{(A.2)}
$$
Denote $-\frac 14 \Delta(f)$ by $c$. Thus, $f= -c\sum_{i\leq 4}u_i\xi_i+f_0$,
where $\Delta(f_0)=0$. By $(2.2')$ then, $\varphi_i^1(u)=2c$ and
$D=Le_{-c\sum{u_i\xi_i}+f_0}+\sum\varphi_i\degree\partial\xi_i+2c\sum\xi_i
\partial\xi_i=Le_{f_0}+\sum\varphi_i\degree\partial \xi_i + cZ$.
Replace $f$ with $f_0$. Then $\Delta(f)=0$ and we have
$$
\pderf{\varphi_i\degree}{u_{j}}
-\pderf{\varphi_j\degree}{u_{i}}=-2\lambda\frac{\partial^2 f}{\partial
{\xi_{l}}\partial {\xi_{k}}}.\eqno{(2.4'')} 
$$
If $\deg_{\xi}f<2$, then the rhs of $(2.4'')$ is equal to 0. So due to Remark
we can take $\varphi_i\degree=0$. In this case
$$
D=Le_f+cZ=D_f+cZ.\eqno{(A.3)}
$$

Let $\deg_\xi f=2$. As $\Delta(f)=0$, we can set $g=\Delta^{-1}(f)$. Then $\deg_\xi g=3$
and $f=\Delta(g)$. For the role of functions
$\varphi_i\degree$ satisfying equation $(2.4'')$ we can take
$$
\varphi_i\degree= 2\lambda\frac{\partial^3 g}{\partial {\xi_{j}}\partial
{\xi_{k}}\partial {\xi_{l}}}, \text{ where }\;  (i, j, k, l)\in A_4.
$$

We get:
$$
D_f=\Le_{f}+2\lambda\sum_{(i, j, k, l)\in A_4; \;
1\leq  i\leq  4}\frac{\partial^3 g}{\partial {\xi_{j}}\partial {\xi_{k}}\partial
{\xi_{l}}}\cdot \partial {\xi_{i}}+cZ=D_f+cZ.\eqno{(A.4)}
$$

Let $\deg_\xi f=3$. Let us represent $f$ in the form $f=f_3+f_{<3}$, where
$f_3$ is an homogeneous (wrt the degree in $\xi$) polynomial of degree 3, while
$f_{<3}$ is a polynomial of lesser degree.

Since $\Delta(f)$ only depends on $u$, we see that $\Delta(f_3)=0$ and,
therefore, we can introduce $H=\Delta^{-1}(f_3) = F(u)\xi_1\xi_2\xi_3\xi_4$ for some
function $F(u)$. 

From $(2.4')$ we deduce that
$$
\pderf{\varphi_i^1}{u_{j}}=2\lambda\pderf{F}{u_{j}} \text{ or, with $(2.2')$, }
\varphi_i^1=2\lambda F= -\frac{1}{2}\Delta(f).
$$
Therefore, $\Delta(f)=\Delta(f_{<3})=-4\lambda F$. Set
$\hat f=4\Delta^{-1}(F)$. We obtain that $f=\Delta(H)-\lambda\hat f+g$ for
some function $g$ such that $\Delta(g)=0$ and $\deg_{\xi}g<3$. But we have already
described the solutions for all such $g$. So now we assume $g=0$. In this
case we can take $\varphi_i\degree=0$ (due to Remark). We get:
$$
\begin{matrix}
D=\Le_f+\lambda(-\Le_{\hat f}+2F\sum\xi_i\partial {\xi_{i}})=\\
\Le_{f}+\lambda\left (-\Le_{\hat f}+2\sum\limits_{(i, j, k, l)\in A_4;\; 1\leq i\leq
4}\frac{\partial^3 H}{\partial \xi_{j}\partial \xi_{k}\partial\xi_{l}}\cdot
\pder{\xi_{i}}\right )=D_f.
\end{matrix}\eqno{(A.5)}
$$
Formulas (A.3)-(A.5) prove Theorem 2.3.

\section*{Appendix 2. Proof of headings 2 of Theorems 4.2, 6.2 and 6.5}

\ssbegin{A.2.1}{Lemma} Let $(\fg _{-1}, \fg _{0})_*$ 
be simple; let the $\fg _{0}$-module $\fg _{-1}$ be irreducible.
If $\fh =(\fg _{-1}, \fcg _{0})_*$ is 
also simple, then for every $v\in \fg _{-1}$
there exists an $F\in \fg _{1}$ such that $[v, F]
\not\in \fg _{0}$.

The same applies to $(\fg _{-}, \fg _{0})^{mk}_*$ 
and $\fh =(\fg _{-}, \fc\fg _{0})^{mk}_*$. \end{Lemma}
 
\begin{proof} By simplicity of $(\fc\fg _{0})_*$ (due to Lemma 4.1.1)
we have $[\fh _{-1}, \fh _{1}]=c\fg _{0}$, 
i.e., there exists $v_{0}\in \fg _{-1}, 
F_{0}\in \fg _{1}$ such that $[v_{0}, F_{0}]\not\in \fg _{0}$.
 
Let
$$
V_{1}=\{v_{1}\in \fg _{-1}: [g, v_{1}]=v_{0}\; \text{for some $g\in \fg _{0}\}$.}
$$
Then for any $v_{1}\in V_{1}$ we have
$$
\fg _{0}\not\ni [[g, v_{1}], F_{0}]=
\pm [[g, F_{0}], v_{1}]\pm [g, [v_{1}, F_{0}]],
$$
where the signs are governed by Sign Rule. Therefore, one of the two cases holds:

1) $[v_{1}, F_{0}]\not\in \fg _{0}$, hence, $F=F_{0}$;

2) $[v_{1}, F_{0}]\in \fg _{0}$.

In case 2) we have $[g, [v_{1}, F_{0}]]\in \fg _{0}$; hence,
$$
[F, v_{1}]\not\in \fg _{0}\; \text{for $F=[g, F_{0}]$.}
$$
 Similarly, introduce the sequence of spaces
$$
V_{2}=\{v_{2}\in \fg _{-1}: [g, v_{2}]\in V_{1}+<v_{0}>\; 
\text{for some $g\in \fg _{0}\}$, etc.}
$$
By irreducibility of $\fg _{0}$-action on $\fg _{-1}$ 
for every $v\in \fg _{-1}$
there exists an $n$ such that $v\in V_{n}$ and, therefore, $F=F_{n}$, 
where $[v, F_{n}]\not\in \fg _{0}$. 

The arguments for depth 2 are literally the same. \end{proof}
 
\ssbegin{A.2.2}{Corollary} Let $\fg _{-1}=\Pi 
(\Lambda (n)/\Cee1), \fg _{0}=\fvect (0|n)$, i.e.,
$\fg _*=(\fg _{-1}, \fg _0)_*=\fle (n; n)$. Then $(\fg _{-1}, \fc(\fg _{0}))_*$ 
is not simple for $n>3$.
\end{Corollary}
 
Hereafter in this Appendix we often abuse the notations and denote the elements by their
generating functions.

\begin{proof} By Lemma A.2.1 the simplicity of $\fg _*$ implies that for
any $v\in \fg _{-1}$ there exists $F\in \fg _{-1}$ 
such that $[v, F]\not\in \fg _{0}$.
 
Take $v=\xi _{1}\ldots
\xi _{n}$; let $d$ be the central element of 
$\fc(\fg _{0})$ normalized so that $\ad~  d|_{\fg _{-1}}=-\text{ id}$. 
Let $F\in \fg _{1}$ be such that
$$
[v, F]=d+g, \; \text{ for}\; g\in \fg _{0}.
$$
Then
$$
\pm [v, [F, v_{1}]]\overset{\text{by Jacobi id.}}=[[v, F],
v_{1}]=(d+g)v_{1}=-v_{1}+gv_{1}. \eqno{(A.2.0)} 
$$
In other words, $g_{1}=[F, v_{1}]$ maps $v$ to 
$-v_{1}+gv_{1}$ up to a sign.
But in the $\fg _{0}$-module considered the element $v$ can only be
mapped into a function of degree $\ge n-1.$
 
Hence, $gv_{1}=v_{1}+\varphi (v_{1})$, where $\deg 
\varphi \ge n-1, $ for any $v_{1}$ of degree $<n-1$. 
Consequently, the projection $g_0$ of $g$ on the zeroth component of $\fvect(0|n)$ with
respect to the standard $\Zee$-grading, i.e., on $\fgl(n)$, satisfies the
condition
$$
g_{0}|_{\Span(v_{1}:\, \deg v_{1}<n-1)}=\id.
$$
But in $\fvect (0|n)$ the dimension of the maximal torus
$\Span(\varepsilon _{i}\partial _{i}: 1\leq i\leq n)$ is equal to $n$ and there 
is no operator whose restrictions
to the spaces of homogenuous functions in $\xi $ of at least
two distinct degrees are scalar operators.
 
Since $n-2\ge 2$ for $n>3$, the Lie superalgebra 
$(\fg _{-1}, \fc\fg _{0})_*$ is not simple.
\end{proof}

\ssbegin{A.2.3}{Corollary} Let 
$\fg _{-}=\left\{\begin{matrix}
\fab(2^{n-1}) &\text{for $n$ odd}\\
\fhei(2^{n-1}|2^{n-1})&\text{for $n$ even}.
\end{matrix}\right .$ The Lie superalgebra $(\fg _{-}, \fcvect (0|n))^{mk}_*$ is not simple
for $n>3$.
\end{Corollary}
 
Proof follows the lines of the proof of Corollary A.2.1
with the correction that (A.2.0) is now true not for all
$v_{1}\in \fg _{-1}$ but only for those which satisfy 
$[v_{1}, v]=0.$ Such
elements $v_{1}$ are represented by functions $f\in \Lambda (n)$ such that
$0<\deg f\le n-1.$ There are $\ge 2$ distinct
degrees which satisfy this inequality for $n>3.$ \qed
 
\ssbegin{A.2.4}{Corollary} Let $\fg _{-}=\left\{\begin{matrix}
\fab(2^{n-1}-1) &\text{for $n$ odd}\\
\fhei(2^{n-1}|2^{n-1}-2)&\text{for $n$ even}.
\end{matrix}\right .$ Then $\fg=(\fg _{-}, \fsvect_{a, b}
(0|n))_*^{mk}$ is not a simple Lie superalgebra if either $n>4$ or $n=4$ and $(a, b)\not
\in\Cee(3, 4)$.
\end{Corollary}

Proof is obtained by a slight modification of the proof of Corollary A.2.2. As $v$
we now take $\xi_1\dots\xi_{n-1}\in\fg_{-1}$; let $F$ be such that $[v,
F]=ax+bd+g$, where $g\in\fsvect(0|n)$. Then 
$$
[[F, v_1], v]=\pm[[v, F], v_1]=(ak-b)v_1+gv_1\eqno{(A.2.1)}
$$
for any monomial $v_1\in\fg_{-1}$ of degree $k$ and distinct from $\xi_n$. Since
every element from $\fg_0$ lowers the degree of any monomial not more than
by 1, we see that the projection $g_0$ of $g$ on $\fsvect(0|n)_0$
satisfies the relation
$$
g_0v_1=(b-ak)v_1\eqno{(A.2.2)}
$$
for any monomial $v_1\in\fg_{-1}$ of degree $k<n-2$ and distinct from $\xi_n$.
In particular, for $n>4$ this means that $g_0$ acts on $\Span(\xi_1, \dots
,\xi_{n-1})$ by multiplication by $b-a$ and on $\Lambda^2(\xi)$ by multiplication
by $b-2a$. Hence, $g_0=0$, i.e., $a=b=0$.

In (A.2.2) $k<n-2$. So if $n=4$, then $k=1$. The component $g_0$ is defined
by its action on $\xi_1$, ..., $\xi_n$. But (A.2.2) gives the action of $g_0$
only on $\xi_1$, ..., $\xi_{n-1}$. Its action on $\xi_n$ can be arbitrary  with
only one condition: $g_0\in \fsvect(0|4)$; this is what (A.2.3) means:
$$
g_0(\xi_1\xi_4)=-2(b-a)\xi_1\xi_4+c_1\xi_1\xi_2+c_2\xi_1\xi_3.\eqno{(A.2.3)}
$$
         
Look at formula (A.2.1) with $v=\xi_1\xi_2\xi_3$ and $v_1=\xi_1\xi_4$.
It means that $ad[F, v_1]$ (which is an element from $\fsvect(0|4)\supplus\Cee(ax+bd)$)
sends $\xi_1\xi_2\xi_3$ to $(2a-b)\xi_1\xi_4 + g(\xi_1\xi_4)$. Since no
vector field can send $v$ to $v_1$, we deduce that $g_0(v_1)$ must compensate
$(2a-b)\xi_1\xi_4$. But from formula (A.2.3) we derive that $b-2a=-2b+2a$,
implying $3b=4a$. \qed

Due to Lemmas 4.1, 6.1, Corollaries A.2.2--A.2.4 are equivalent to the headings 2) of
Theorems 4.2, 6.2 and 6.5, respectively.

\section*{Appendix 3. Proof of simplicity of the Lie superalgebras \protect $\fg =
(\fab(4), \fcvect (0|3))^{m}_*$ and $\fg =(\fhei(8|6), 
\fsvect_{3, 4}(0|3))^{k}_*$}

$$
\begin{matrix}\text{Due to Lemma 6.1, to prove the simplicity of $\fg$ it suffices to
exibit an element
$\hat F\in\fg_1$}\\ 
\text{such that
$[\fg_{-1}, \hat F]$ is not entirely contained in $\fvect(0|3)$ and
$\fsvect(0|3)$, respectively.}\end{matrix}\eqno{(A.3.1)}
$$

\ssec{A.3.1. Simplicity of $(\fab(4), \fcvect (0|3))^{m}_*$} First, let us show how
to embed $\fg =\fcvect (0|3)^{m}_*$ into $\fm (4)$. We consider $\fm (4)$
as preserving the Pfaff equation given by the
form $\alpha_{0}=d\tau+\sum^{3}_{i=0}(\eta _{i}du_{i}+u_{i}d\eta _{i})$. Denote
the basis elements of $\fg $ as follows
$$
\renewcommand{\arraystretch}{1.3}
\begin{tabular}{|c|c|c|}
\hline
$\fg _{-2}$&A basis of $\fg _{-1}=\Pi (\Vol^\frac 12)$& 
notations of the corr. functions that generate $\fg _{-1}\subset\fm(4)$\\ 
\hline
&$\xi _{1}\xi _{2}\xi _{3}$&$\eta _{0}$\\ 
$M_1$ &$\xi _{2}\xi _{3}$, $\xi _{3}\xi _{1}$, 
$\xi _{1}\xi _{2}$& $u_{1}$, $u_{2}$, $u_{3}$\\ 
& $\xi _{1}$, $\xi _{2}$, $\xi _{3}$ &$\eta _{1}$, 
$\eta _{2}$, $\eta _{3}$\\ 
& $1$ &$ u_{0}$\\
\hline
\end{tabular}
$$
 
The following is an explicit realization of the
embedding $i: \fg _{0}=\fvect (0|3)
\longrightarrow \fle (3)$. We only indicate the generating functions of the image:
$$
\renewcommand{\arraystretch}{1.3}
\begin{tabular}{|c|c|c|}
\hline
$\deg D$&$D\in \fvect (0|3)$&$i(D)$\\ 
\hline
$-1$&$\partial _{1}$, $\partial _{2}$, $\partial _{3}$ 
& $-u_{0}u_{1}+\eta _{2}\eta _{3}$, $-u_{0}u_{2}+\eta _{3}\eta _{1}$,
$-u_{0}u_{1}+\eta _{1}\eta _{2}$\\ 
\hline
$0$&$\Div D=0:\quad \xi _{i}\partial _{j}$& 
$-u_{i}\eta_{j}$ for $i\neq j$; $i, j>0$\\ 
$0$&$\xi _{1}\partial _{1}$& 
$\frac 12(u_{0}\eta_{0}-u_{1}\eta_{1}+u_{2}\eta_{2}+u_{3}\eta_{3})$\\ 
$0$&$\xi _{2}\partial _{2}$& 
$\frac 12(u_{0}\eta_{0}+u_{1}\eta_{1}-u_{2}\eta_{2}+u_{3}\eta_{3})$\\ 
$0$&$\xi _{3}\partial _{3}$& 
$\frac 12(u_{0}\eta_{0}+u_{1}\eta_{1}+u_{2}\eta_{2}-u_{3}\eta_{3})$\\ 
\hline
$1$&$\xi_{2}\xi_{3}\partial _{1}$, $\xi_{3}\xi_{1}\partial _{2}$, 
$\xi_{1}\xi_{2}\partial _{3}$& $-\frac 12u^{2}_{1}$, $-\frac 12u^{2}_{2}$,
$-\frac 12u^{2}_{3}$\\ 
$1$&$\xi_{1}(\xi_{2}\partial _{2}- \xi_{3}\partial _{3})$, 
$\xi_{2}(\xi_{3}\partial _{3}- \xi_{1}\partial _{1})$, \;
$\xi_{3}(\xi_{1}\partial _{1}- \xi_{2}\partial _{2})$&$-u_{2}u_{3}$, \;
$-u_{1}u_{3}$,\; $-u_{1}u_{2}$\\ 
$1$&$\xi_{1}(\xi_{2}\partial _{2}+\xi_{3}\partial _{3})$, \;
$\xi_{2}(\xi_{3}\partial _{3}+ \xi_{1}\partial _{1})$, \;
$\xi_{3}(\xi_{1}\partial _{1}+\xi_{2}\partial _{2})$& $\eta _{0}\eta _{1}$,\;
$\eta _{0}\eta _{2}$, \; $\eta _{0}\eta _{3}$\\ 
\hline
$2$&$\xi _{1}\xi _{2}\xi _{3}\partial _{i}$&$-\frac 12u_{i}\eta_{0}$\\
\hline
\end{tabular}
$$ 
 
To check condition (A.3.1), take
$$
\hat F=M_F, \; \text{where}\; F=2\tau u_{1}-2\eta _{0}\eta _{2}\eta _{3}+u_{0}^2\eta _{1}.
$$
Then the brackets with $\fg _{-1}$ are
$$
\begin{matrix}
\{F, u_{0}\}_{m.b.}&=&-2u_{0}u_{1}+2\eta _{2}\eta _{3}, \\ 
\{F, u_{1}\}_{m.b.}&=&-3u_{1}^2,\\ 
\{F, u_{2}\}_{m.b.}&=&-2u_{1}u_{2}-2\eta _{0}\eta _{3};\end{matrix}
\qquad
\begin{matrix}
\{F, \eta _{1}\}_{m.b.}&=&2\tau;\\ 
\{F, \eta _{i}\}_{m.b.}&=&-2u_{1}\eta _{i} \; (i=0, 2, 3); \\ 
\{F, u_{3}\}_{m.b.}&=&-2u_{1}u_{3}+2\eta _{0}\eta _{2}.\end{matrix}\eqno{(A.3.2)}
$$
We get $M_\tau$, while the remaining elements in the rhs of (A.3.2) lie in $\fvect(0|3)$.
\qed

\ssec{A.3.2. Simplicity of $(\fhei(8|6), \fsvect_{3, 4}(0|4))_*^{k}$} First, let us
show how to embed $\fsvect_{3, 4}(0|4)_*^{k}$ into $\fk (9|6)$. We realize
$\fk (9|6)$ as preserving the Pfaff equation given by the
form $\alpha_{1}=dt-\sum\limits^{}_{i\le 4}(p_{i}dq_{i}-dq_{i}p_{i})-
\sum\limits^{}_{j\le 3}(\eta _{j}d\xi _{j}+\xi _{j}d\eta _{j})$. Let us redenote
the basis elements of $\fg_{-1}$:
$$
\renewcommand{\arraystretch}{1.3}
\begin{tabular}{|c|c|}
\hline
A basis of $\fg _{-1}$ &notations of the corr. functions that generate $\fg
_{-1}\subset\fk (9|6)$\\ 
\hline
$\xi _{1}$, $\xi _{2}$, $\xi _{3}$, $\xi _{4}$& 
$p_{1}$, $p_{2}$, $p_{3}$, $p_{4}$\\ 
$\xi _{1}\xi _{2}$, $\xi _{1}\xi _{3}$, $\xi _{1}\xi _{4}$& 
$\eta _{1}$, $\eta _{2}$, $\eta _{3}$\\ 
$-\xi _{3}\xi _{4}$, $\xi _{2}\xi _{4}$, $-\xi _{2}\xi _{3}$&
$\zeta _{1}$, $\zeta _{2}$, $\zeta _{3}$\\ 
$\xi _{2}\xi _{3}\xi _{4}$, $ -\xi _{1}\xi _{3}\xi _{4}$, 
$\xi _{1}\xi _{2}\xi _{4}$, $-\xi _{1}\xi _{2}\xi _{3}$& 
$q_{1}$, $q_{2}$, $q_{3}$, $q_{4}$\\
\hline
\end{tabular}
$$
The following is an explicit realization of the
embedding $i: \fg _{0}=\fsvect (0|3)
\longrightarrow \fh(8|6)_0$. We only indicate the generating functions of the image.
For $D\in\fsvect(0|3)$ we have

$\underline{\deg D=-1:}$ 
$$
\begin{matrix}
\partial _{\xi _{1}}&\mapsto \zeta _{1}p_{2}+
\zeta _{2}p_{3}+\zeta _{3}p_{4} \\ 
\partial _{\xi _{2}}&\mapsto -\zeta _{1}p_{1}+
\eta _{2}p_{4}-\eta _{3}p_{3} \\ 
\partial _{\xi _{3}}&\mapsto -\zeta _{2}p_{1}+
\eta _{3}p_{2}-\eta _{1}p_{4} \\ 
\partial _{\xi _{4}}&\mapsto -\zeta _{3}p_{1}-
\eta _{2}p_{2}+\eta _{1}p_{3} \end{matrix}
$$
$\underline{\deg D=0:}$
$$
\begin{matrix}
\xi _{1}\partial _{2} & \mapsto -p_{1}q_{2}-\eta _{2}\eta _{3} \\ 
\xi _{2}\partial _{1} & \mapsto -p_{2}q_{1}+\zeta _{2}\zeta _{3} \\ 
\xi _{1}\partial _{3} & \mapsto -p_{1}q_{3}+\eta _{1}\eta _{3} \\ 
\xi _{3}\partial _{1} & \mapsto -p_{3}q_{1}-\zeta _{1}\zeta _{3} \\ 
\xi _{1}\partial _{4} & \mapsto -p_{1}q_{4}-\eta _{1}\eta _{2} \\ 
\xi _{4}\partial _{1} & \mapsto -p_{4}q_{1}+\zeta _{1}\zeta
_{2}\end{matrix}
\qquad\begin{matrix}
\xi _{2}\partial _{3} & \mapsto -p_{2}q_{3}+\eta _{1}\zeta _{2} \\ 
\xi _{3}\partial _{2} & \mapsto -p_{3}q_{2}+\eta _{2}\zeta _{1} \\ 
\xi _{2}\partial _{4} & \mapsto -p_{2}q_{4}+\eta _{1}\zeta _{3} \\ 
\xi _{4}\partial _{2} & \mapsto -p_{4}q_{2}+\eta _{3}\zeta _{1} \\ 
\xi _{3}\partial _{4} & \mapsto -p_{3}q_{4}+\eta _{2}\zeta _{3} \\ 
\xi _{4}\partial _{3} & \mapsto -p_{4}q_{3}+\eta _{3}\zeta
_{2}\end{matrix}
$$
$$ 
\begin{matrix}
\xi _{1}\partial _{1}-\xi _{2}\partial _{2} & 
\mapsto -p_{1}q_{1}+p_{2}q_{2}+
\eta _{2}\zeta _{2}+\eta _{3}\zeta _{3}\\
\xi _{2}\partial _{2}-\xi _{3}\partial _{3} & 
\mapsto -p_{2}q_{2}+p_{3}q_{3}+
\eta _{1}\zeta _{1}-\eta _{2}\zeta _{2}\\
\xi _{3}\partial _{3}-\xi _{4}\partial _{4} & 
\mapsto -p_{3}q_{3}+p_{4}q_{4}+
\eta _{2}\zeta _{2}-\eta _{3}\zeta _{3}\\ 
\sum\xi _{i}\partial _{i} & 
\mapsto -\sum p_{i}q_{i}-2t\hfil\end{matrix}\eqno{(A.3.3)}
$$
$\underline{\deg D=1:}$

$$
\xi _{1}\xi _{2}\partial _{3}\mapsto -\eta _{1}q_{3}, \text{ etc.}
$$ 
$$
\begin{matrix}
\xi _{1}\xi _{2}\partial _{1}+\xi _{2}\xi _{3}
\partial _{3} & \mapsto -q_{1}\eta _{1}+q_{3}\zeta _{3} \\ 
\xi _{1}\xi _{2}\partial _{1}+\xi _{2}\xi _{4}
\partial _{4} & \mapsto -q_{1}\eta _{1}-q_{4}\zeta _{2} \\ 
\xi _{1}\xi _{3}\partial _{1}+\xi _{3}\xi _{2}
\partial _{2} & \mapsto -q_{1}\eta _{2}-q_{2}\zeta _{3} \\ 
\xi _{1}\xi _{3}\partial _{1}+\xi _{3}\xi _{4}
\partial _{4} & \mapsto -q_{1}\eta _{2}+q_{4}\zeta _{1}\end{matrix}
\qquad\begin{matrix}
\xi _{1}\xi _{4}\partial _{1}+\xi _{4}\xi _{2}
\partial _{2} & \mapsto -q_{1}\eta _{3}+q_{2}\zeta _{2} \\ 
\xi _{1}\xi _{4}\partial _{1}+\xi _{4}\xi _{3}
\partial _{3} & \mapsto -q_{1}\eta _{3}-q_{3}\zeta _{1} \\ 
\xi _{1}\xi _{2}\partial _{2}-\xi _{1}\xi _{3}
\partial _{3} & \mapsto -q_{2}\eta _{1}+q_{3}\eta _{2} \\ 
\xi _{1}\xi _{2}\partial _{2}-\xi _{1}\xi _{4}
\partial _{4} & \mapsto -q_{2}\eta _{1}+q_{4}\eta _{3}\end{matrix}
$$
$\underline{\deg D=2:}$ The image under $i$ is generated by
$q_{i}q_{j}$ for any $1\leq i, j\leq 4$.

Now, set 
$$
\begin{matrix}
x_0=K_{-\sum p_{i}q_{i}-2t}, &x_1=K_{-p_{1}q_{1}+p_{2}q_{2}+
\eta _{2}\zeta _{2}+\eta _{3}\zeta _{3}},\\
x_2=K_{-p_{2}q_{2}+p_{3}q_{3}+
\eta _{1}\zeta _{1}-\eta _{2}\zeta _{2}},&x_3=K_{-p_{3}q_{3}+p_{4}q_{4}+
\eta _{2}\zeta _{2}-\eta _{3}\zeta _{3}},
\end{matrix}
$$ 
see (A.3.3). Set:
$$
f=t+\sum^{}_{i\le 3}p_{i}q_{i}+3p_{4}q_{4}+
\eta _{1}\zeta _{1}+\eta _{2}\zeta _{2}-\eta _{3}\zeta _{3}.
$$
Then
$$
K_f=\frac 12x_1+x_2+
\frac 32x_3-\frac 32x_0-2K_t\in\fsvect(0|4)\supplus \Cee(3x_0+4K_t).
$$
To check the condition (A.3), take
$$
\hat F=K_F, \; \text{where}\; F=tp_4+p_4(\sum^{}_{i\leq 4}p_{i}q_{i}+
\eta _{1}\zeta _{1}+\eta _{2}\zeta _{2}-\eta _{3}\zeta _{3})-2\zeta _1\zeta
_2p_1+2\zeta _1\eta _3p_2+2\zeta _2\eta _3p_3. 
$$
The commutators of $F$ with $\fk_{-1}(9|6)$ are of the form: 
$$
\begin{matrix}
\{q_{i}, F\}_{k.b.}&=q_{i}\pderf{F}{t}+
\pderf{F}{p_{i}}; \\ 
\{p_{i}, F\}_{k.b.}&=p_{i}\pderf{F}{t}-
\pderf{F}{q_{i}}; \end{matrix}
\qquad\begin{matrix}
\{\eta _{i}, F\}_{k.b.}&=\eta _{i}\pderf{F}{t}-
\pderf{F}{\zeta _{i}}; \\ 
\{\zeta _{i}, F\}_{k.b.}&=\zeta _{i}\pderf{F}{t}-
\pderf{F}{\eta _{i}}.\end{matrix}
$$
Hence,
$$ 
\begin{matrix}
\{q_4, F\}_{k.b.}&=&f; \\ 
\{\eta_1, F\}_{k.b.}&=&2(\eta _{1}p_{4}+\zeta _{2}p_1-\eta _{3}p_2)\mapsto
-2\partial_3;\\ 
\{\eta_2, F\}_{k.b.}&=&2(\eta _{2}p_{4}-\zeta _{1}p_1-\eta _{3}p_3)\mapsto
2\partial_2;\\ 
\{\eta_3, F\}_{k.b.}&=&\{\zeta _{1}, F\}_{k.b.}= \{\zeta _{2}, F\}_{k.b.}=0;\\ 
\{\zeta_3, F\}_{k.b.}&=&2(\zeta _{3}p_{4}+\zeta _{1}p_2+\zeta _{2}p_3)\mapsto
2\partial_1;\\ 
\{q_1, F\}_{k.b.}&=&2(q_1p_{4}-\zeta _{1}\zeta_2\mapsto
-2\xi_4\partial_1;\\ 
\{q_2, F\}_{k.b.}&=&2(q_2p_{4}+\zeta _{1}\eta_3\mapsto
-2\xi_4\partial_2;\\ 
\{q_3, F\}_{k.b.}&=&2(q_3p_{4}+\zeta _{2}\eta_3\mapsto
-2\xi_4\partial_3;\\ 
\{p_i, F\}_{k.b.}&=&0\quad\text{ for }i=1, 2, 3, 4.\end{matrix}
$$
So we get $K_f$, while the remaining brackets lie in $\fsvect(0|3)$. \qed

\end{document}